\documentclass[11pt]{article}
\pdfoutput=1
\usepackage{jcappub}
\usepackage{mathtools}
\usepackage{microtype}
\usepackage[utf8]{inputenc}
\usepackage[english]{babel}
\usepackage[dvipsnames]{xcolor}
\usepackage{amsmath,amssymb,amsbsy,amstext,amsthm}
\usepackage[colorlinks=true]{hyperref}
\usepackage{microtype}
\usepackage{graphicx}
\usepackage{amsfonts}
\usepackage{upgreek}
\usepackage{exscale,relsize}
\usepackage[makeroom]{cancel}
\usepackage{soul}
\usepackage{float}
\usepackage{empheq}
\usepackage[most]{tcolorbox}
\RequirePackage{color}
\usepackage{feynmp}
\usepackage{ifpdf}
\ifpdf
\fi

\usepackage{colortbl}
\definecolor{green2}{cmyk}{0, 1, 0.5, 0}
\definecolor{lightgreen}{cmyk}{0.2, 0, 0.2, 0.2}
\definecolor{dred}{rgb}{0.9,0.2,0.5}
\definecolor{dred2}{cmyk}{0.1,0.7,0.1,0.3}
\definecolor{lightgray2}{cmyk}{0.4,0.4,0,0.8}
\definecolor{black}{cmyk}{1.0,1.0,1.0,1.0}

\definecolor{verde}{rgb}{0,0.5,0}

\AtBeginDocument{
 \hypersetup{
 urlcolor=NavyBlue,
 citecolor=NavyBlue,
 linkcolor=BrickRed,
 }
}

\allowdisplaybreaks[1]

\usepackage{colortbl}

\makeatletter
\newlength{\apb@width}
\newcommand{\autoparbox}[2][c]{\settowidth{\apb@width}{#2}\parbox[#1]{\apb@width}{#2}}

\makeatother

\numberwithin{equation}{section}

\def\beq{\begin{equation}}
\def\eeq{\end{equation}}

\def\bea{\begin{eqnarray}}
\def\eea{\end{eqnarray}}

\def\ni{\noindent}

\def\0{{\boldsymbol 0}}

\def\x{{\boldsymbol{x}}}

\newtcbox{\mymath}[1][]{%
    nobeforeafter, math upper, tcbox raise base,
    enhanced, colframe=gray!30!gray,
    colback=gray!10, boxrule=0.5pt,
    #1}

\usepackage{setspace} 

\begin{document}

\title{Pure Chromo-Natural Inflation: Signatures of Particle Production from Weak to Strong Backreaction}

\author[\diamondsuit]{Ema Dimastrogiovanni,}
\author[\spadesuit]{Matteo Fasiello,}
\author[\spadesuit,1]{Alexandros Papageorgiou\note{corresponding author},}
\author[\spadesuit]{and Crist\'{o}bal Zenteno Gatica}

\affiliation[\diamondsuit]{Van Swinderen Institute for Particle Physics and Gravity, University of Groningen, Nijenborgh 4, 9747 AG Groningen, The Netherlands}
\affiliation[\spadesuit]{Instituto de F\'{i}sica T\'{e}orica UAM-CSIC, c/ Nicol\'{a}s Cabrera 13-15, 28049, Madrid, Spain}

\emailAdd{e.dimastrogiovanni@rug.nl}
\emailAdd{matteo.fasiello@csic.es}
\emailAdd{papageorgiou.hep@gmail.com}
\emailAdd{cristobal.zenteno@ift.csic.es}

\vspace{1.2cm}

\abstract{
\ni We consider, in the context of axion-inflation, the \textit{Pure Natural Inflation} (PNI) model coupled with an SU(2) gauge sector via a Chern-Simons term. As the axion rolls down its potential, it dissipates energy in the gauge sector thus sourcing fluctuations of scalar and tensor degrees of freedom therein. Gauge field fluctuations will, in turn, feed primordial gravitational waves as well as curvature perturbations. Remarkably, we can use upcoming cosmological probes to test this mechanism across a vast range of scales, from the CMB to laser interferometers. 
Due to their flat plateau at large field values, we find that PNI potentials fare better vis-\'{a}-vis CMB observations than the conventional sinusoidal potential of chromo-natural inflation (CNI). We show that, even when the dynamics begin in the weak backreaction regime, the rolling of the axion leads to a build-up of the gauge-quanta production, invariably triggering the strong backreaction of the gauge sector tensors on the background dynamics. This transition results in the copious production of both scalar and tensor perturbations, which we study in detail.  The gravitational wave signatures include a rich peak structure with a characteristic scale-dependent chirality, a compelling target for future gravitational wave detectors. Additionally, the peak in scalar perturbations may lead to the formation of primordial black holes, potentially accounting for a significant fraction of the observed dark matter abundance.}

\maketitle

\section{Introduction}

The existence of viable alternatives \cite{Brandenberger:2009jq} notwithstanding, the possibility of an accelerated expansion in the very early universe, inflation, stands as one of the main pillars of the cosmological standard model \cite{Lyth:1998xn}. The inflationary paradigm elegantly solves some of the key puzzles of the hot big-bang model and explains how the observed CMB temperature anisotropies originate from quantum fluctuations. Amplified by the expansion, these fluctuations provide the seeds for the growth of structure and may act as a mechanism for primordial black hole (PBH) production thus supplying a compelling candidate for dark matter \cite{Bird:2016dcv}. \\ \indent 
The energy scales at play during the early acceleration, likely many order of magnitude above those accessible via particle colliders, make inflationary observables a unique portal into (beyond-) Standard Model physics with the intriguing prospect of capturing glimpses of quantum gravity \cite{Baumann:2014nda}.
The burgeoning number of (soon-to-be) available cosmological probes has emphatically improved our ability to test inflation \cite{SimonsObservatory:2018koc,CMB-S4:2016ple,CMB-S4:2022ght,Amendola:2016saw,LSSTScience:2009jmu}. Cosmic microwave background experiments shed light on the dynamics that took place about sixty e-folds before the acceleration ends. Pulsar timing arrays and gravitational detectors probe, through the gravitational signal \cite{EPTA:2023xxk,NANOGrav:2023gor,SKA:2018ckk,LIGOScientific:2016wof,LIGOScientific:2021aug,LISACosmologyWorkingGroup:2022jok,Maggiore:2019uih,Reitze:2019iox,Kawamura:2020pcg}, the  intermediate and late stages of inflation accessing a largely unexplored  realm, ripe with information on the particle content of the very early universe.

Despite the tremendous success of inflation vis \`{a} vis CMB observations, ours is still a broad-brush description of the acceleration mechanism. One could not be faulted for noting that large-scale observations have been cementing the importance of the inflationary paradigm, but have so far stopped short of identifying a preferred (class of) inflationary model(s). The microphysics of inflation, its (self)interactions and particle content remain elusive, as attested by the fact that the single \textit{vs} multi-field dichotomy has yet to be settled. Fortunately, an unprecedented array of cosmological probes will be launched in the coming decade(s); when combined, these hold the potential to deliver a ``high-definition'' picture of inflation. Qualitatively crucial threshold will soon be crossed in terms of the tensor-to-scalar ratio ($\sigma_r\sim 10^{-3}$ via CMB-S4, LiteBIRD), and the non-linear parameter $f_{\rm NL}$ ($\sigma_{f_{\rm NL}}\sim \textit{\rm a\;few}$ via LSS survey,  $\sigma_{f_{\rm NL}}\lesssim 0.1$ via 21-cm line measurements \cite{Munoz:2015eqa}). There is more: intermediate and small-scale probes such as PTA and gravitational wave detectors will test for a signal associated uniquely to multi-field realizations of inflation. From the unparalleled amount and quality of the cosmological data we will collect emerges the rather realistic prospect of being able, in the not too distant future, to pinpoint a preferred class of inflationary models, if not a specific model itself.

In addition to the undergoing monumental effort on the side of observations, there is much progress to be made on inflationary model-building and, more generally, top-down approaches to inflation. First, one would like to build models that directly tackle the $\eta$-problem: the large quantum corrections the inflaton mass would receive in the absence of a symmetry protecting its necessarily small value. Perhaps the most compelling proposal in this context is that of natural inflation \cite{Freese:1990rb,Adams:1992bn}: the (approximate) shift-symmetry characterizing this construction ensures a technically natural small inflaton mass.  Although   ruled out by observations\footnote{At least in its original disguise, see \cite{Freese:2014nla}.}, natural inflation serves as an inspiration for any axion-inflation model with naturalness ambitions, including the very mechanism we will study in this work. 

The ever-increasing constraining power of CMB measurements has very often led towards considering multi-field realizations of axion inflation models. The typically steep potential of the axion can be made effectively flatter by, for example, an added source of friction. As a welcome byproduct, the axion decay constant in these constructions may well be sub-Planckian. Interestingly, both the multi-field content of these models and the sub-Planckian size of $f$ happen to also be the typical outcome of string theory realizations of inflation. In selecting efficient criteria for model selection, it has proven very useful to pair up viability constraints with the requirement that inflationary models be embeddable\footnote{See e.g. \cite{DallAgata:2018ybl,Holland:2020jdh,Dimastrogiovanni:2023juq} for embeddings of axion gauge field models analogous to the ones we will discuss in this work.} in UV complete theories such as string theory.   

In this work we are after a natural multi-field model of axion inflation that can be embedded in higher dimensional constructions. Our starting point is the model of pure natural inflation (PNI) \cite{Nomura:2017ehb}. We refer the interested reader to the original literature \cite{Nomura:2017ehb,Nomura:2017zqj} for further details on how such mechanism is natural from the top-down perspective. To the simple ingredients of \cite{Nomura:2017ehb}, namely a single axion-inflaton with a well-motivated potential driving acceleration, we add an SU(2) gauge sector via Chern-Simons (CS) coupling. Note that it is  natural to envisage the presence of such dimension-5 operators from the EFT point of view and that gauge fields are easily accommodated in string theory. The (approximate) shift symmetry is clearly preserved by the CS term up to a total derivative. We term this model \textit{pure chromo-natural inflation} (PCNI) in that it has the same field content as chromo-natural inflation \cite{Adshead:2012kp,Dimastrogiovanni:2012ew,Dimastrogiovanni:2012st,Maleknejad:2012fw,Adshead:2013nka,Maleknejad:2013npa,Maleknejad:2014wsa,Adshead:2016omu,Maleknejad:2016dci,Maleknejad:2018nxz,Papageorgiou:2018rfx,Lozanov:2018kpk,Agrawal:2018mkd,DallAgata:2018ybl,Mirzagholi:2019jeb,Fujita:2022fff,Dimastrogiovanni:2023oid,Murata:2024urv,Michelotti:2024bbc} but shares the axion-inflaton potential with PNI.

At the background level, coupling the axion with a non-Abelian sector makes the homogeneous and isotropic FRW solution possible even in the presence of a non-zero gauge field VEV \cite{Galtsov:1991un,Maleknejad:2011jw}. Such configuration has two desirable properties. First, it is an attractor  as  anisotropic solutions isotropize within a few e-folds \cite{Maleknejad:2011jr,Wolfson:2020fqz}. Secondly, as shown in \cite{Domcke:2018rvv,Domcke:2019lxq}, if one starts out with effectively multiple Abelian fields (with vanishing VEV) the  chromo configuration with a non-zero gauge field VEV will emerge dynamically as a result of the inflationary evolution. As the inflaton field rolls down its potential, dumping energy into the gauge sector provides some extra\footnote{In addition to the ever-present Hubble friction, $3H\dot{\phi}$ .} friction. The six degrees of freedom in the gauge sector react to this dumping: their fluctuations  significantly affect observables, including the gravitational wave spectrum \cite{Papageorgiou:2018rfx,Papageorgiou:2019ecb,Badger:2024ekb}. As a result of the parity-breaking nature of the CS coupling, the two polarization of tensor modes in the gauge sector 
behave rather differently. Most strikingly, one polarization undergoes a finite instability and is thus enhanced. Given its linear coupling to gravity, this gives rise to rich and intriguing gravitational wave  signatures, including a chiral signal soon to be tested by upcoming probes \cite{Seto:2007tn,Smith:2016jqs,Domcke:2019zls,Unal:2023srk}.

The parameter that best describes the energy dissipation across the two sectors is the particle production\footnote{The particles in question here being the gauge field quanta $A^{a}_{\mu}\,$.} parameter $\xi=\lambda \dot{\phi}/(2fH)$. Its definition makes it clear that the effect of the gauge tensor on observables cannot but increase as the inflaton velocity $\dot{\phi}$ grows during the evolution. This is indeed the case but, crucially, only up to the onset of the so-called strong backreaction regime, that is the point where (tensor) gauge field fluctuations become so amplified that their effect on the background equations can no longer be neglected. In this work we follow the evolution of the system well-within the strong backreaction regime. Strong backreaction effects have been extensively studied in the Abelian case, both analytically \cite{Anber:2009ua,Peloso:2022ovc,vonEckardstein:2023gwk} and on the lattice \cite{Caravano:2022epk,Figueroa:2023oxc,Caravano:2024xsb,Figueroa:2024rkr,Sharma:2024nfu}. 

The non-Abelian configuration has been tackled only very recently \cite{Iarygina:2023mtj,Dimastrogiovanni:2024xvc}, and under the assumption of an homogeneous inflaton background: these results are rather interesting. In \cite{Iarygina:2023mtj} an attractor regime was first identified where the gauge field VEV stops growing and asymptotes to a  relatively small negative value. The time dependence of the gauge field VEV profile ensures that its value crosses  an interval which  corresponds to a possibly perilous instability in the scalar sector \cite{Dimastrogiovanni:2012ew}. In \cite{Dimastrogiovanni:2024xvc} we explored such instability in depth, drawing two main lessons. First and foremost, depending on the hierarchy between certain key parameters, the instability can be tamed and harnessed to deliver large scalar fluctuations and even generate primordial black holes. Secondly, the work in \cite{Dimastrogiovanni:2024xvc} has shown how strong backreaction dynamics is largely insensitive to the global properties of the potential. As long as the scene is set to trigger strong backreaction, the salient parts of the gauge VEV evolution, including the crossing of the instability band and approaching the attractor solution, will all take place within a number of e-folds corresponding to a very small inflaton field excursion. This is in fact what makes the dynamics in question blind to global profile of the potential.

The choice of the potential remains instead paramount\footnote{It has indeed proven rather difficult to identify minimal models satisfying CMB constraints while displaying interesting, testable, signatures. This has lead to intense research activity in theories equipped with spectator axion sectors (see e.g.\cite{Barnaby:2012xt,Mukohyama:2014gba,Namba:2015gja,Garcia-Bellido:2016dkw} and \cite{Obata:2016tmo,Dimastrogiovanni:2016fuu,Agrawal:2017awz,Thorne:2017jft,Fujita:2017jwq,Agrawal:2018gzp,Dimastrogiovanni:2018xnn,Fujita:2018vmv,Papageorgiou:2019ecb,Mirzagholi:2019jeb,Mirzagholi:2020irt,Wolfson:2020fqz,Iarygina:2021bxq,Fujita:2021flu,Kakizaki:2021mgj,Ishiwata:2021yne,Bagherian:2022mau,Iarygina:2023mtj,Putti:2024uyr,Brandenburg:2024awd}), a well-motivated possibility in the context of the so-called string-axiverse \cite{Arvanitaki:2009fg,Acharya:2010zx,Cicoli:2012sz,Demirtas:2018akl,Demirtas:2021gsq, DAmico:2021vka,DAmico:2021fhz,Dimastrogiovanni:2023juq}.} to satisfy CMB bounds. It is indeed the preference for a  natural potential together with the need to recover the appropriate scalar power spectrum (and tilt) at CMB scales that leads us to the choice of the pure natural inflation  framework. We will show how the strong backreaction dynamics has the same general features uncovered in \cite{Dimastrogiovanni:2024xvc}, and acts as a mechanism for scalar induced gravitational waves and, possibly, also  PBH production. 

Equipped with the PNI potential, we couple the axion to an SU(2) gauge sector and study in detail the essentially inevitable transition between weak and strong backreaction regime  during inflation. We uncover striking features in a gravitational wave spectrum including a three-peak structure which is robust to significant changes in the parameters. A peak-like structure also arises in the scalar power spectrum and we discuss under what conditions it supports significant PBH production. For a fiducial set of parameters, we show how the GW signal is detectable by LISA and identify smoking-gun signatures related to the chiral properties of the GW spectrum. 

 This paper is organized as follows. In \textit{Section}~\ref{sec:chromo-regimes} we give an overview of the various regimes of CNI-like models paying particular attention to analytic results in the weak and strong backreaction attractors. In \textit{Section}~\ref{sec:pure-chromo} we specify the choice of the potential, making the model concrete. In \textit{Section}~\ref{sec:pheno} we explore in detail the phenomenology of our setup and calculate the observables corresponding to a set of representative fiducial parameters. Finally, in \textit{Section} \ref{sec:conclusion} we draw our conclusions and comment on future work. The main text is supplemented by several appendices that contain lengthy but important derivations omitted from the main body.

\section{The various regimes of Chromo-Natural Inflation}
\label{sec:chromo-regimes}

As briefly outlined in the Introduction, the CNI model exhibits a rich and interesting structure characterized by a weak and strong backreaction regime. We devote this section to presenting the model and reviewing the basic features of each regime. An underlying assumption is the existence of a non-zero homogeneous gauge field background\footnote{The line element is given by $ds^2=-dt^2+a(t)^2 d\vec{x}^2=a(\tau)^2\left(-d\tau^2+d\vec{x}^2\right)$ in physical and conformal time respectively. Indexes $a={1,2,3}$ and $i={1,2,3}$ are defined in SU(2) and SO(3) spaces respectively while index $0$ is reserved for time.}

\begin{equation}
    A^a_0=0\;\;\;,\;\;\; A^a_i=\delta^a_i a(t) Q(t)\;,
\end{equation}
where $Q(t)$ is a time-dependent function regulating the strength of the gauge field background. The action of our model takes the form

\begin{equation}
    S=\int {\rm d}^4 x\sqrt{-g}\left[\frac{M_p^2}{2} R-\frac{1}{2}\left(\partial\chi\right)^2-V(\chi)-\frac{1}{4}F^a_{\;\mu\nu} F^{a\,\mu\nu}+\frac{\lambda \chi}{4 f} F^a_{\;\mu\nu}\tilde{F}^{a\,\mu\nu}\right]\;.
    \label{eq:CNIL}
\end{equation}

where $f$ is the axion decay constant and $F^a_{\; \mu\nu}\equiv\partial_\mu A_\mu -\partial_\nu A_\mu -g \,\epsilon^{abc} A^b_\mu A^c_{\nu}$ is the gauge field strength. The gauge field self-coupling parameter is $g$ and the Chern-Simons term includes the dual field strength tensor defined as $\tilde{F}^{a\, \mu\nu}\equiv\frac{\epsilon^{\mu\nu\rho\sigma}}{2\sqrt{-g}}F^a_{\;\rho\sigma}$. Throughout this work we define the number of e-folds as $N\equiv \ln a$ and normalize $a_{\rm CMB}=1$.\\
Strictly speaking, the Lagrangian in Eq.(\ref{eq:CNIL}) describes CNI only upon specifying the potential \bea
V(\chi) =\mu^4 \left(1+\cos\frac{\chi}{f}\right)\; .
\eea
When the potential in Eq.~(\ref{eq:PNIP}) is instead  the one employed,  Eq.(\ref{eq:CNIL}) will correspond to PCNI. In what follows, we shall remain agnostic about the potential for as long as possible in order to highlight the dynamics common to both models.
\subsection{Background equations of motion with backreaction}

The equations of motion follow directly from the variation of the action, 

\begin{align}
    \ddot{\chi}+3H\dot{\chi}&+V_{\chi}+\frac{3 g \lambda}{f}Q^{2}\left(\dot{Q}+H Q\right)+\mathcal{T}^{\chi}_{BR}=0\,,\label{eq:eqchi}\\
    \ddot{Q}+3H\dot{Q}&+\left(\dot{H}+2H^{2} \right)Q+gQ^{2}\left(2gQ-\frac{\lambda \dot{\chi}}{f} \right)+\mathcal{T}^{Q}_{BR}=0\,,\label{eq:eqQ}
\end{align}

where $V_\chi$ denotes the derivative of the potential with respect to the axion $\chi$. In the above equations the backreaction of  perturbations is denoted as ${T}^{\chi}_{BR}$ and ${\cal T}^{Q}_{BR}$ for the axion and the gauge field equation of motion respectively. The first and second Friedmann equations read 
\begin{align}
    3 H^{2} M_{p}^{2}=&
    \frac{\dot{\chi}^{2}}{2}+V(\chi)+\frac{3}{2}\Big[\left(\dot{Q}+HQ \right)^{2}+g^{2}Q^{4}\Big]+\rho_t\,,\label{eq:H}\\
    -2M_{p}^{2}\dot{H}=&
    \dot{\chi}^{2}+2\left[\left(\dot{Q}+HQ \right)^{2}+g^{2}Q^{4}\right]+Q \,{\cal T}^Q_{BR}+\frac{4}{3}\rho_t\,.\label{eq:Hdot}
\end{align}

\ni with $\rho_t$  the total energy of the fluctuations. It is worth commenting at this stage on the presence of the backreaction term ${\cal T}_{BR}^Q$ in the equation for $\dot{H}$. The second field equation contains a term proportional to the second derivative of the gauge field background. The latter does not appear in Eq.~(\ref{eq:Hdot}) because we have used the equation of motion for $Q$ to replace it on the right-hand side of (\ref{eq:Hdot}) with the backreaction term ${\cal T}^Q_{BR}$ .

The second Friedmann equation can be rearranged in the following way
\begin{align}
    \epsilon_H\equiv -\frac{\dot{H}}{H^2}=\epsilon_\chi+\epsilon_B+\epsilon_E+\epsilon_{{\cal T}_Q}+\epsilon_{\rho_t}\;,
\end{align}
with
\begin{align}
    \epsilon_\chi\equiv \frac{\dot{\chi}^2}{2M_p^2 H^2}\,,&\;\;\epsilon_B\equiv \frac{g^2Q^4}{M_p^2 H^2}\,,\;\;\epsilon_E\equiv \frac{\left(\dot{Q}+H Q\right)^2}{M_p^2 H^2}\,,\;\;\epsilon_{{\cal T}^Q_{BR}}\equiv&\frac{Q}{2 M_p^2 H^2}{\cal T}^Q_{BR}\,,\;\;\epsilon_{\rho_t}\equiv\frac{2}{3 M_p^2 H^2}\rho_t\, .
\label{eq:slow}
\end{align}

\subsection{Tensor fluctuations}

As is well known, this model contains two tensor degrees of freedom in the gauge sector that are linearly coupled with tensor fluctuations of the metric. These gauge tensor modes play a crucial role in the overall phenomenology of the model and, in particular, in determining the onset of the strong backreaction regime.

The gauge field contains two traceless transverse degrees of freedom which can be decomposed (one may choose, without loss of generality, the momentum orientation to be along the z axis) as $\delta A^1_\mu=a(0,t_+,t_\times,0)$ and $\delta A^2_{\mu}=a(0,t_\times, -t_+,0)$. Changing  basis and working in momentum space\footnote{We move to momentum space with the Fourier convention $\delta(t,\vec{x})=\int \frac{{\rm d}^3 k}{\left(2\pi\right)^{3/2}}{\rm e}^{i\vec{k}\cdot\vec{k}}\delta\left(t,\vec{k}\right)$. The circular polarizations of the tensor perturbations (gauge and metric) are defined as $t_{L,R}\equiv a (t_+\pm i\,t_{\times})$, $h_{L,R} \equiv (a M_{p}/2)(h_+ \pm i\, h_{\times})$. We use the "hat" notation to denote canonically quantized modes.}, we can write the equations of motion for the circular polarizations of the tensors in the gauge sector and for gravitational waves as

\begin{align}
\ddot{\hat{t}}_{R,L}&+H\,\dot{\hat{t}}_{R,L}+\left[\frac{k^2}{a^2}\pm\frac{k}{a}\left(2 g Q + \frac{\lambda \dot{\chi}}{f}\right)+\frac{g Q\lambda \dot{\chi}}{f}\right]\hat{t}_{R,L} =\nonumber\\
&\hspace{0cm}
+\frac{2}{M_p}\left(\dot{Q}+H Q\right)\dot{\hat{h}}_{R,L} -\frac{2}{M_p}\Bigg[\mp \frac{k}{a}g Q^2+ H\left(\dot{Q} + H Q\right)+  g Q^2\left(g Q-\frac{\lambda \dot{\chi}}{f}\right)+{\cal T}^Q_{BR}\Bigg]\hat{h}_{R,L}\label{eq:tmodes}  
\end{align}
\begin{align}
\ddot{\hat{h}}_{R,L}+H\,\dot{\hat{h}}_{R,L} + \Bigg[\frac{k^2}{a^2}-2H^2-\frac{\left(\dot{Q}+HQ\right)^2}{M_p^2}&+\frac{3 g^2 Q^4}{M_p^2}+\frac{\dot{\chi}^2}{2M_p^2}-\frac{3 Q {\cal{T}}^Q_{BR}}{2 M_p^2}\Bigg]\hat{h}_{R,L} =\nonumber\\
&\hspace{0cm}
-\frac{2}{M_p}\left(\dot{Q}+H Q\right)\dot{\hat{t}}_{R,L}+\frac{2g Q^2}{M_p}\left(\pm\frac{k}{a}+g Q\right) \hat{t}_{R,L}\label{eq:hmodes}    
\end{align}

Accounting for the backreaction of the tensor perturbations on the background equations of motion is essential for a consistent treatment of the model dynamics.

\begin{equation}
\begin{array}{rcl}
     \mathcal{T}^{\chi}_{BR}&=&
     -\frac{\lambda}{2 f a^3} \int \frac{d^3k}{(2 \pi)^3}\frac{d}{dt}\left[ 
      \left(a g \, Q + k \right)\left\vert\hat{t}_R\right\vert^2 +
     \left(a g \, Q - k \right)\left\vert\hat{t}_L\right\vert^2
     \right]
     \\
     \mathcal{T}^{Q}_{BR}&=&  
     \frac{g}{3a^2}  \int \frac{d^3k}{(2 \pi)^3}
     \left[
     \left(\frac{\lambda \dot{\chi}}{2 f}+ \frac{k}{a}\right)\left\vert \hat{t}_R \right\vert^2 + 
     \left(\frac{\lambda \dot{\chi}}{2 f}- \frac{k}{a}\right)\left\vert \hat{t}_L\right\vert^2
     \right]
     \\
     \rho_{\tau}&=&\frac{1}{2a^2}\int 
     \frac{d^3k}{(2 \pi)^3}
     \left[
     \left\vert \dot{\hat{t}}_R\right\vert^2 +\left(\frac{k^2}{a^2} + 
     2 g Q \frac{k}{a}\right) \left\vert \hat{t}_R\right\vert^2 +
     \left\vert \dot{\hat{t}}_L\right\vert^2 +\left(\frac{k^2}{a^2}-
     2  g Q \frac{k}{a}\right) \left\vert \hat{t}_L\right\vert^2
     \right]
\end{array}
\label{eq:back1}
\end{equation}

Note that terms suppressed by inverse powers of the Planck mass have been omitted in Eq.~\ref{eq:back1}.
Using the standard notation

\begin{equation}
    m_Q\equiv \frac{g Q}{H} \;\;\;,\;\;\; \Lambda\equiv \frac{\lambda Q}{f }\;\;\;,\;\;\;\xi\equiv \frac{\lambda \dot{\chi}}{2H f} \;,
\end{equation}

\noindent one can re-write the homogenous part of the gauge tensor perturbations using the variable $x\equiv -k \tau$.

\begin{equation}
    \hat{t}''_{R,L}+\left[1+\frac{2}{x^2}\left(m_Q\xi\pm x(m_Q+\xi)\right)\right]\hat{t}_{R,L}=0\label{eq:dispersion}
\end{equation}

Eq.~(\ref{eq:dispersion}) shows how, depending on the relationship between parameters $m_Q$ and $\xi$, it is possible for the two polarizations of the gauge field tensor fluctuations to become ``tachyonic" inside the horizon. This naturally increases the occupation number of gauge tensor modes leading to an enhanced gravitational wave production and potentially causing strong backreaction on the equations of motion for the zero mode. Since parameters $m_Q$ and $\xi$ control directly the amount of particle production, we will refer to them as ``particle production" parameters.
In the next section we will take a closer look at the conditions required to arrive at such a tachyonic enhancement both in the weak and strong backreaction regimes.

\subsection{The weak backreaction regime}
\label{sec:low}

The weak backreaction regime of CNI has been extensively studied \cite{} so that we can limit our analysis to the key notions and formulas. In such regime the slow roll conditions $\ddot{Q}\simeq\ddot{\chi}\simeq 0$ are in place. Studies have so far centered on the configuration corresponding to ``strong friction'', which is made possible thanks to the magnetic helicity term arising from the gauge field background\footnote{One can arrive at such a configuration by requiring that $\Lambda\gg 2$ and $\Lambda \gg \sqrt{3}/m_Q$.}.
In such configuration, the Hubble friction term is negligible and the gauge field background is, to a good approximation, frozen at the bottom of its effective potential ($\dot{Q}=0$). 

In this regime the backreaction is negligible (${\cal T}^Q_{BR}={\cal T}^\chi_{BR}=0$). The motion of the background fields is governed by the system 

\begin{align}
    &V_{\chi}+\frac{3 g \lambda H}{f}Q^{3}\simeq 0\,,\\
    &2H^{2}Q+gQ^{2}\left(2gQ-\frac{\lambda \dot{\chi}}{f} \right)\simeq 0\,,
\end{align}

The terms above can be rearranged into the ``master" formulas that describe the weak backreaction regime in CNI:

\begin{equation}
    Q\simeq\left(\frac{-f V_\chi}{3 g \lambda H}\right)^{1/3}\;\;\;,\;\;\;\xi\simeq m_Q+\frac{1}{m_Q}\; .
    \label{eq:CNI-attractor}
\end{equation}
In light of the second expression in Eq.~(\ref{eq:CNI-attractor}), we will sometimes refer to $m_Q$ (and not just to the standard $\xi$) as the particle production parameter. It should be clear from the context to which parameter we are referring to at any given point.  Assuming, without loss of generality, a positive particle production parameter $\xi$ and focusing on the phenomenologically more interesting region where $\xi\gg 1$, will automatically fix the sign of the  $m_Q$ parameter in the weak backreaction regime. Let us then study the dispersion relation of the $\hat{t}_{R,L}$ fluctuations to understand whether they undergo a tachyonic instability. Using formula (\ref{eq:CNI-attractor}) in Eq.~(\ref{eq:dispersion}) and setting the dispersion relation to zero, we find that in this regime the $t_R$ polarization can never become tachyonic at any point (be it inside or outside the horizon). On the other hand, the $t_L$ polarization becomes tachyonic in the interval between

\begin{equation}
    x_{L,{\rm inst},\pm}=\frac{1}{m_Q}+2m_Q\pm\sqrt{\frac{1}{m_Q^2}+2\left(1+m_Q^2\right)}>1\;.
    \label{eq:instthres}
\end{equation}

This signals a strong tachyonic enhancement of the $t_L$ mode which starts inside the horizon at the time $x_{L,{\rm inst},+}$ and ends while still inside the horizon at point $x_{L,{\rm inst},-}$. Note that the solutions are such that the tachyonic instability is always guaranteed to switch off before horizon crossing in the weak regime of CNI. For the phenomenologically relevant parameter space\footnote{That is, the region that delivers, in the weak backreaction regime, the measured value of scalar density perturbations at CMB scales.}, this strong tachyonic instability invariably leads the system into the strong backreaction regime. The minimum value of the particle production parameter $m_Q$ for a given coupling strength $g$ that is required to enter the strong backreaction regime has been analytically estimated in the literature to be (to lowest order in slow-roll):

\begin{equation}
    g\leq \left(\frac{24 \pi^2}{2.3\, {\rm e}^{3.9 m_Q}}\frac{1}{1+\frac{1}{m_Q^2}}\right)^{1/2}
    \label{estimateB}
\end{equation}

\begin{figure}
\begin{center}
\includegraphics[scale=0.7]{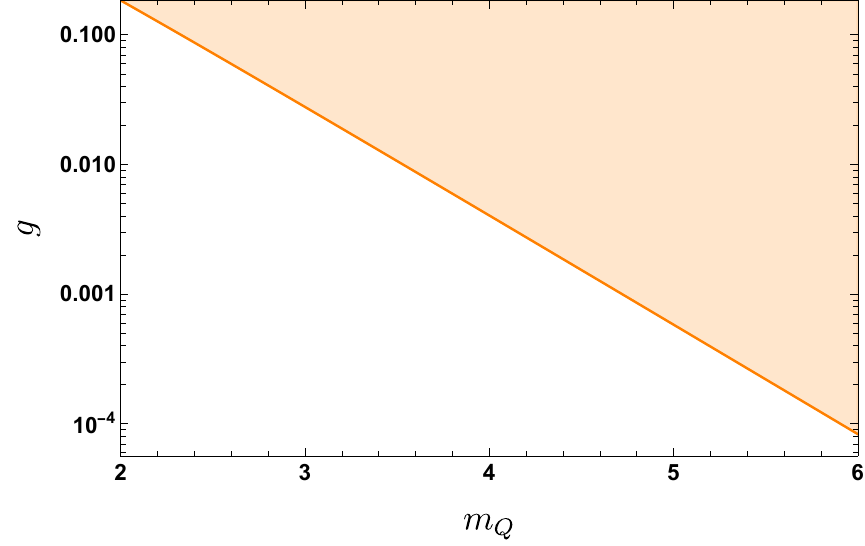}
\end{center}
\vspace*{-5mm}\caption{The orange shaded area displays the region of parameter space whose analysis requires we take backreaction into account.}
\label{fig:strong-backreaction-1}
\end{figure}

Numerical results confirm Eq.~(\ref{estimateB}) is a good approximation. We plot the region of parameter space accessible in the weak backreaction regime (WBR) in Fig.~\ref{fig:strong-backreaction-1}. The typical evolution in weak backreaction is characterized by a small value of $m_Q$ that is continuously growing as the axion moves further down the potential. Requiring that $P_\zeta$ take the observed value at CMB scales and making sure to stay within the allowed range of scalar non-Gaussianity generally restricts the particle production parameter to satisfy $m_{Q,{\rm CMB}}\leq 2.5$ (See App.~\ref{app:density_perturbation} for more details). It is universally the case that in the WBR $\epsilon_H\simeq \epsilon_B$ since the magnetic helicity is the main source of friction responsible for slowing down the motion of the axion. We can estimate the final value of the particle production parameter by taking the ratio of the slow roll parameters at the moment of horizon crossing of CMB modes and at the end of inflation

\begin{equation}
    \frac{\epsilon_{H,{\rm CMB}}}{\epsilon_{H, {\rm END}}}\simeq\frac{H^2_{\rm CMB} m^4_{Q,{\rm CMB}}}{H^2_{\rm END} m^4_{Q,{\rm END}}} \;\;\;\Rightarrow\;\;\; m_{Q,{\rm END}}\simeq\left(\frac{H_{\rm CMB}}{H_{\rm END}}\right)^{1/2}\frac{m_{Q,{\rm CMB}}}{\epsilon^{1/4}_{H,{\rm CMB}}}\;.
    \label{eq:mQestimate}
\end{equation}

Even under the most conservative of assumptions for which $H_{\rm CMB}\simeq H_{\rm END}$ and $\epsilon_{H,{\rm CMB}}\simeq 10^{-3}$ we find that 

\begin{equation}
    m_{Q,{\rm END}} \gtrsim 10\;.
\end{equation}

Comparing this number to the bound displayed in Fig.~\ref{fig:strong-backreaction-1} it is readily apparent that entering the strong backreaction regime before the end of inflation is unavoidable unless one chooses an extremely small value of the gauge coupling constant $g$. It turns out that such values of $g$ are hardly compatible with the observed value of $n_s$, unless the potential is severely fine-tuned. We therefore conclude that the most general scenarios in CNI will exhibit a weak backreaction regime followed by a strong backreaction one. A transition from weak to strong backreaction typically comes with an explosive production of perturbations (both scalar and tensor) which may leave highly distinct signatures at intermediate and small scales. We explore the dynamics of the transition  in the next subsection.

\subsection{The transition from weak to strong backreaction}
\label{sec:transition}

The transition is a highly non-linear process during which there is copious particle production. We devote this subsection to describing, at the qualitative level, the various phases of the transition from one regime to the other and give a bird's-eye view of the various particle production mechanisms  expected to be relevant to the phenomenology of the model.

\begin{figure}
\begin{center}
\includegraphics[scale=0.5]{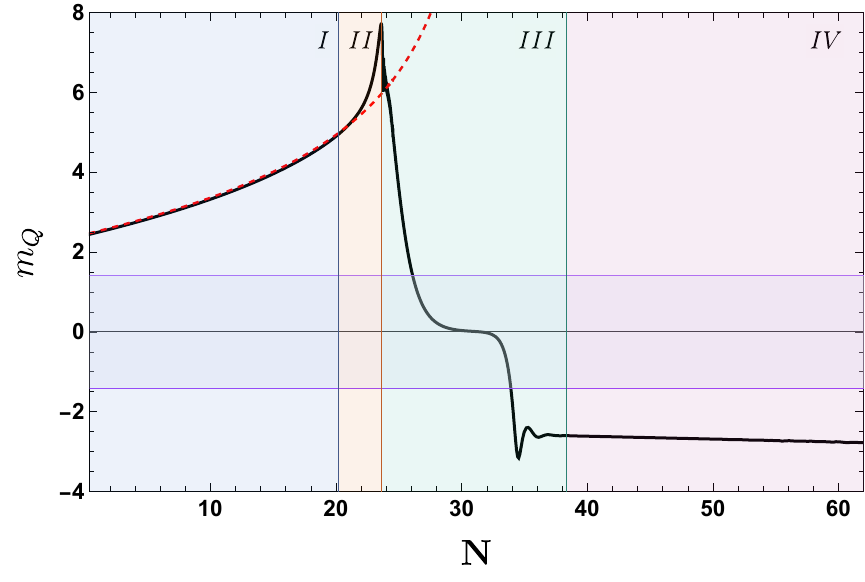}
\includegraphics[scale=0.5]{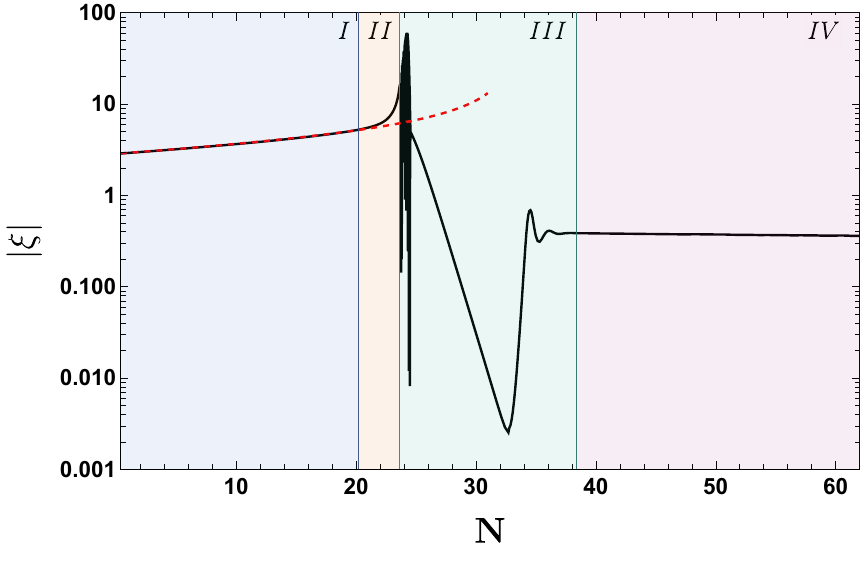}
\end{center}
\vspace*{-5mm}\caption{A typical example of the various phases that take place in CNI whereby the dynamics starts in the weak backreaction regime and ends in the strong one. A detailed description of the phases is in the main text and the parameters chosen for this run are found in Sec.~\ref{sec:pheno}, formula (\ref{eq:fid}). The horizontal axis represents e-folds,  from CMB to the end of inflation. We observe a qualitatively similar behavior for the parameters $m_Q,\xi$. The black solid lines are the case with backreaction while the red dashed line disregards the effects of backreaction.}
\label{fig:transition}
\end{figure}

Fig.~\ref{fig:transition}  displays a typical example of the non-trivial dynamics occurring during the transition. We identify the following phases:

\begin{itemize}
    \item Phase I: is the initial \textit{weak backreaction} phase during which the CMB modes leave the horizon. During this period, the particle production parameter $m_Q$ increases monotonically until it reaches approximately the value obtained by means of the analytical estimate for the onset of strong backreaction  (for a given parameter $g$, this can be obtained from Fig.~\ref{fig:strong-backreaction-1}). The parameters $\xi, m_Q$ are related as in (\ref{eq:CNI-attractor}).
    \item Phase II: During this short phase, the particle production parameters increase with respect to the evolution in the absence of backreaction.  This means that the axion is briefly experiencing a short \textit{kick} or acceleration down the potential due to the backreaction. This phase always lasts for no more than a few e-folds.
    \item Phase III: This is the \textit{transition phase} during which the particle production parameter $\xi$ decreases, crosses zero and eventually settles to a negative quasi-constant value. 
    \item Phase IV: This is the \textit{strong backreaction} phase in which the slow roll of the axion and the gauge field background $Q$ is supported by backreaction terms in the background equations of motion. At this stage the contribution of, respectively, $\mathcal{T}^{\chi}$ and $\mathcal{T}^{Q}$ in Eqs.~(\ref{eq:eqchi}) and (\ref{eq:eqQ}) is comparable to those of other leading terms.
\end{itemize}

There are at least three complementary effects that contribute towards interesting signatures such as a GW signal at intermediate/small scales and PBH production.
The first effect is the direct production of gravitational waves which takes place at the end of Phase I and during Phase II. During this period, the particle production parameters $m_Q$ and $\xi$ are at their maximum values so that the ensuing tachyonic production of gauge tensor perturbations will significantly  source GWs via the linear coupling described by Eqs.~(\ref{eq:tmodes})-(\ref{eq:hmodes}).

Another effect is the strong tachyonic production of scalar perturbations occurring while the $m_Q$ parameter crosses the $m_Q< |\sqrt{2}|$ divide in Phase III. This instability was originally analyzed in \cite{Dimastrogiovanni:2012ew,Adshead:2013nka}; we review the physics relevant to the present analysis in App.~\ref{app:scalar}. The resulting scalar perturbations may generate PBH and, correspondingly, scalar induced GWs once they re-enter the horizon during radiation domination. Finally, during Phase III the axion-inflaton dramatically slows down as clear from the large decrease in the value of $\xi\propto \dot{\chi}$. This will necessarily translate into an enhancement of the density perturbations since the two are related via $\zeta\simeq - \delta\chi\, H/\dot{\chi}$ (see App.~\ref{app:density_perturbation}). The findings we present in detail in the Phenomenology section will reflect the more qualitative discussion above.

\subsection{The strong backreaction regime}
\label{sec:strong}

The strong backreaction regime was originally studied in \cite{Iarygina:2023mtj} where  a well-defined late-time attractor was identified. We summarize some key  points in this section. The strong backreaction attractor is characterized by the relation
\begin{equation}\label{NewAttEq}
    m_Q \xi = -1\,,
\end{equation}
between $m_Q$ and $\xi$.

\begin{figure}
\begin{center}
\includegraphics[scale=0.7]{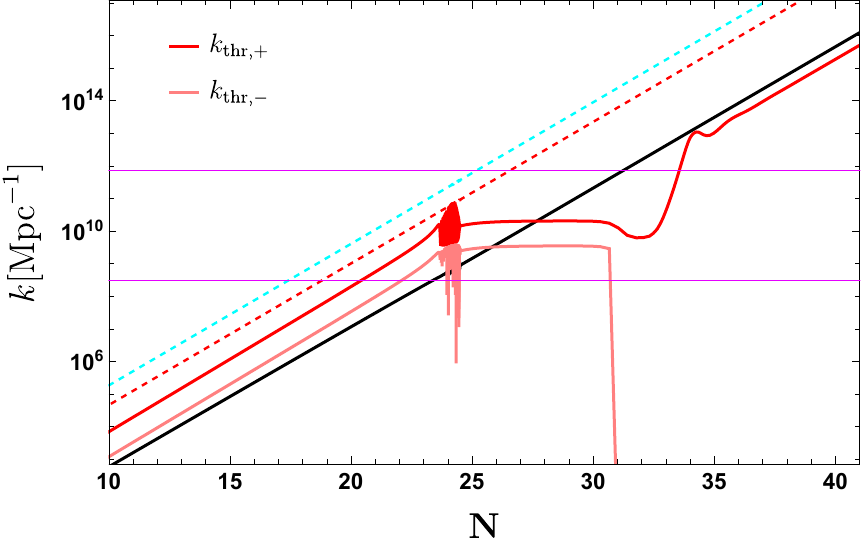}
\end{center}
\vspace*{-5mm}\caption{Evolution of the instability band for the same example as in Fig.~\ref{fig:transition}. The black solid line shows the inverse comoving Hubble radius $a H$, the red and pink solid lines display the instability thresholds defined in eq.~(\ref{eq:instthres}). Note how the instability band moves from inside to outside the horizon as explained in the main text. The red dashed line displays the time dependent cutoff in momentum space we set for the backreaction integrals while the cyan dashed line displays the moment beyond which we evolve each mode from their vacuum configuration (well-enough before their respective onset of instability). Finally, the two purple horizontal lines show the relevant momentum interval that has to be considered to get the full backreaction effect on the background at every moment in time.}
\label{fig:kthr}
\end{figure}

This is a direct result of the non-trivial dynamics between the instability in the tensor perturbations and the backreaction terms entering the equations of motion. Specifically, in the strong backreaction regime, the backreaction integrals are dominated by a narrow range of modes which are the last modes to have become tachyonic inside the horizon in the weak backreaction regime (see Fig.~\ref{fig:kthr}). It is straightforward to see this from the dispersion relation in the new attractor regime. Using Eq.~(\ref{NewAttEq}) in Eq.~(\ref{eq:dispersion}) one obtains 

\begin{equation}
    x_{L,\pm}=-\frac{1}{m_Q}+m_Q\pm\sqrt{\frac{1}{m_Q^2}+m_Q^2}\;.
\end{equation}

The above solutions  correspond to one which is formally negative and hence unphysical and another threshold which is always smaller than one for $m_Q<-\sqrt{2}$. This implies that the once strong instability in the left-handed mode which took place inside the horizon is now effectively ``tamed" in the strong backreaction attractor as long as $m_Q$ takes negative values. As a result, the profile of the $t_L$ perturbations is controlled by the last modes which became tachyonic in the weak backreaction regime through the function $c(k)$ defined below. These dominant modes are now far outside the horizon and their evolution is universal and characterized by 

\begin{equation}\label{LeftModesNAtt}
    x \, \hat{t}_L \sim \frac{c(k)}{\sqrt{2 k}}    
\end{equation}

This universal evolution allows one to simpify the backreaction terms 

\begin{equation}
\begin{array}{rcl}
{\cal T}^\chi_{BR} &\simeq& -\frac{3 H g \lambda Q}{2 f a^2}\int \frac{d^3 k}{\left(2\pi\right)^3}\left\vert\hat{t}_L\right\vert^2\\
{\cal T}^Q_{BR} &\simeq& \frac{\lambda g\dot{\chi}}{6 f  a^2}\int \frac{d^3 k}{\left(2\pi\right)^3}\left\vert \hat{t}_L\right\vert^2
\end{array}
\label{eq:backsimp}
\end{equation}

and one can do without calculating the integrals themselves since the key quantity is the ratio 

\begin{equation}
    \frac{{\cal T}^Q_{BR}}{{\cal T}^\chi_{BR} } \simeq \frac{2}{9}\frac{H f}{\lambda g Q^2}\;.
\end{equation}

The equations of motion for the background fields $\chi$ and $Q$ can now be simplified to a one-to-one relation between the slope of the axion potential and the particle production parameter $m_Q$. 

\begin{equation}\label{PotRelationNA} 
\frac{dV}{d\chi} =
\frac{3 \lambda}{f g^2} H^4 m_Q^3
\left[ 5 + 3 m_Q^2  
+ \frac{2 f^2 g^2}{m_Q^4 H^2 \lambda^2} \right].
\end{equation}

The formulas (\ref{PotRelationNA}) and (\ref{NewAttEq})  describe the strong backreaction attractor and their role is equivalent to the set in Eq.~(\ref{eq:CNI-attractor}) for the weak backreaction regime. Similar to the WBR, we can write an approximate relation for the slow-roll parameter that is useful in determining the end of inflation. In this case, there are two leading contributions to $\epsilon_H$, 

\begin{equation}\label{epsH-StrongBCK}
    \epsilon_H \approx \epsilon_B + \epsilon_{\rho_t} = \frac{H^2m_Q^2}{g^2M_{p}^2} \left(4+3 m_Q^2 \right)\;  .
\end{equation}

\subsection{Challenges associated to ending inflation in the strong backreaction regime}
\label{sec:Challenges}

The discovery of a well-defined, analytic, strong backreaction attractor makes it easier to aim for a thorough description of the transition from the weak to the strong backreaction regime. This dynamics comes with its own unique signatures during the transition and one may hope to eventually end inflation in a smooth way.  As it turns out, all of the above is indeed possible for the model in hand, except for a smooth end to inflation. The end of inflation faces two important challenges, which we describe in detail in this subsection.

\subsubsection{Instability of the right-handed polarization}
\label{sec:RHmodes}

\begin{figure}
\begin{center}
\includegraphics[scale=0.7]{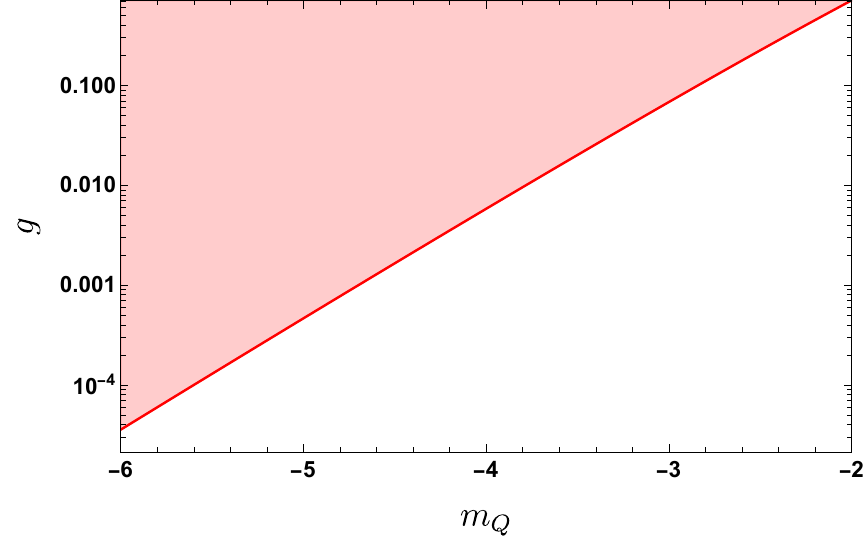}
\end{center}
\vspace*{-5mm}\caption{The red-shaded area denotes the parameter space for which the backreaction from the production of the right-handed tensor mode of the gauge field can no longer be disregarded.}
\label{fig:strong-backreaction-2}
\end{figure}

As outlined in the previous section, once the system enters the strong backreaction regime, the particle parameter $m_Q$ takes negative values. As a result, it is in principle possible for a novel tachyonic instability to emerge in the right-handed polarization of the gauge tensor perturbation which we have so far neglected as is normally done in the literature. Implementing the strong backreaction attractor in the dispersion relation (\ref{eq:dispersion}) and setting it to zero we arrive at two solutions
\begin{equation}
    x_{R,{\rm inst},\pm} = \frac{1}{m_Q} - m_Q \pm \sqrt{\frac{1}{m_Q^2} + m_Q^2}\;.
\end{equation}
One solution,  $x_{R,{\rm inst},-}<0$, is negative and therefore inconsequential whereas the other one is positive and signals a switch in the sign of the dispersion relation deep inside the horizon, $x_{R,{\rm inst},+}>1$, for $m_Q<-\sqrt{2}$. As a result, there is a strong tachyonic instability of the ``neglected'' right-handed mode. Now, in complete analogy with the analysis of \cite{Dimastrogiovanni:2016fuu,Maleknejad:2018nxz,Papageorgiou:2019ecb} in the weak backreaction regime, we can require that the backreaction due to the tachyonically enhanced right-handed mode is much smaller than the smallest term in the effective potential of the gauge field background $Q$. This leads to an upper limit on the self-coupling $g$ for a given  $m_Q$
\begin{equation}
    g\leq \left(\frac{38 \pi^2 m_Q^2}{0.05\, e^{-5.5 m_Q }}   \right)^{1/2}\; ,
    \label{eq:SB2}
\end{equation}
which we plot in Fig.~\ref{fig:strong-backreaction-2}. 

Since during the strong backreaction regime $m_Q$ takes negative order one values that tend to increase in magnitude, it is expected that the inequality in Eq.~(\ref{eq:SB2}) may be violated before the end of inflation. This conclusion holds for values of $g\simeq {\cal O}(10^{-4})$ which, as pointed out in the previous section, are necessary to obtain the measured value for  $P_\zeta$ at CMB scales. The possibility of a strong backreaction for the right-handed mode signals the existence of a third ``phase": a new strong backreaction regime sustained by the right-handed mode (rather than the left-handed one). 

Modeling the transition between the two regimes would require describing the simultaneous evolution of both left and right-handed modes. The associated characteristic scales are rather different given that the right-handed modes (those behind the second strong backreaction phase) would cross the horizon many e-folds after the onset of the first strong backreaction regime. Exploring such a multi-scale configuration is beyond the scope of the present work. However, as we will show in the next  subsection there exists a potential challenge to the existence of this second strong backreaction phase.
\subsubsection{Superhorizon decay of the scalar perturbations}
\label{sec:largeeta}

Another challenge to the smooth ending of inflation in the strong backreaction attractor comes in the form of too large an effective mass in the scalar perturbations of the model. Using the conventional notation $\eta_\chi\equiv V''(\chi)/(3 H^2)$ we can write the effective mass of the axion-inflaton perturbation as follows (see App.~\ref{app:scalar}):
\begin{equation}
    m_{\hat{X}}^2(x) = \left[-2  + \varepsilon_H + 3\eta_\chi + x^2 \left(1+\frac{m_Q^2 \Lambda^2}{2m_Q^2+x^2}\right)\right]H^2\;.
\end{equation}
Assuming that inflaton dominates the energy density, the parameter $\eta_\chi \simeq M_p^2 V''(\chi)/V(\chi)$ formally diverges as the axion reaches the minimum of the potential at the origin, in the absence of a cosmological constant. In the conventional single-field slow-roll scenario, inflation ends before the parameter $\eta_\chi$ becomes large and hence the perturbations remain frozen at super horizon scales. However, in the strong backreaction attractor there is so much  friction due to the combined effects of the gauge field background and the perturbations backreaction that inflation persists until the axion field nearly reaches the bottom of the potential. The corresponding  $\eta_\chi$ parameter is very large\footnote{We note here that, depending on the shape of the potential, it is possible for the perturbations to become tachyonically enhanced for a large negative $\eta_\chi$. Our analysis holds independently of the sign of $\eta_\chi$.}.

In order to quantify how large $\eta_\chi$ needs to be to significantly impact the superhorizon evolution of the perturbations, we require that the mass term is important at  horizon crossing ($x\sim 1$) and consider $2m_Q^2\gg 1$. This yields an upper bound, $\eta_\chi\leq \Lambda^2/6$. If the system inflates for values of the parameter exceeding  this bound, all superhorizon perturbations (including CMB modes) will significantly decay due to their large effective mass. 

\begin{figure}
\begin{center}
\includegraphics[scale=0.7]{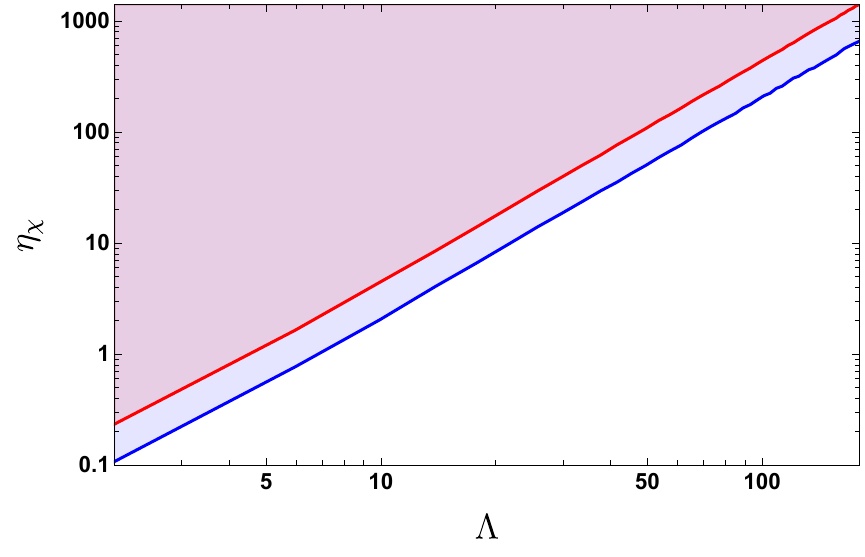}
\end{center}
\vspace*{-5mm}\caption{The blue and red lines correspond to a $10 \% $ and $20\%$ fractional deviation (with respect to the massless case) of the amplitude of a superhorizon mode evaluated 5 e-folds after horizon crossing.}
\label{fig:etL1020Const}
\end{figure}

In practice the above bound , even though easily obtained analytically, is not enough to ensure the freezing of  perturbations at superhorizon scales. We shall instead derive a more conservative bound by solving the system of scalar perturbations in App.~\ref{app:scalar} for several values of $\Lambda$, $\eta_\chi$. We require that the fractional deviation of the axion superhorizon scalar perturbation with respect to the exact massless ($\eta_\chi=0$) case  be less than $20\%$. The result is displayed in Fig.~\ref{fig:etL1020Const}.
In order to avoid the large-$\eta$ suppression, we find the following approximate bound reliable
\begin{equation}
\eta_\chi\lesssim \Lambda^2 \times 10^{-2}\; ,
\label{eq:etalimit}
\end{equation} 
so that for $\Lambda\simeq {\cal O}(1)$ an  $\eta_\chi\sim 1$  makes the scalar perturbations too heavy.  

To compare the bound on $\eta_\chi$ to the expected values of the same parameter in the strong backreaction attractor, we differentiate Eq.~(\ref{PotRelationNA}) with respect to the field value on both sides and use it to express the geometric factor in the strong backreaction regime (the details of the derivation are present in App.~\ref{app:analytics}):
\begin{equation}
    \eta_{\chi,{\rm END}}  \approx \frac{3}{2} m_Q^4 \Lambda^2\, {\cal O}(\epsilon_H)\,.
\label{eq:etaapprox}
\end{equation}
Note that in order to arrive at Eq.~(\ref{eq:etaapprox}) we have also performed a Taylor expansion for large  $m_Q$ and hence it should be understood as an accurate expression near the end of inflation. Comparing the bound in Eq.~(\ref{eq:etalimit}) with Eq.~(\ref{eq:etaapprox}) it becomes clear that a strong backreaction dynamics occurring at any time other than towards the very end of inflation would result in an unacceptably large suppression of scalar perturbations.  Indeed, proceeding along similar lines as in Eq.~(\ref{eq:mQestimate}), we can relate the  parameter $m_Q$  at some time $A$ during inflation in the strong backreaction regime to the its value at a time $B$, which we take here to be the end of inflation. Making use of Eq.~(\ref{epsH-StrongBCK}) 
\begin{equation}
    \frac{\epsilon_{A}}{\epsilon_{B}} =\frac{H_{A}^2 m_{Q,A}^2}{H_{B}^2 m_{Q,B}^2} \frac{(4/3 + m_{Q,A}^2 )}{(4/3 + m_{Q,B}^2) }    
\end{equation}
it is straightforward to see that even under the most generous of conditions for which $H_A=H_B$, $m_{Q,A}=-\sqrt{2}$, $\epsilon_A\simeq 10^{-2}$ and $\epsilon_B\simeq 1$, the parameter at the end of inflation must be at least of order $m_{Q,B}\simeq -5$ .

We conclude that the bound in Eq.~(\ref{eq:etalimit})  will typically be violated well before the end of inflation. 

In closing this chapter, there are several aspects of the analysis that are worth emphasizing. The bound on $\eta_\chi$ in Eq.~(\ref{eq:etalimit}) is going to be violated by the end of inflation in the strong backreaction regime, but it is also expected to be mildly exceeded by the time the right-handed tensor modes back-react on the background equations of motion as laid out in Sec.~\ref{sec:RHmodes}\footnote{Note that Eq.~(\ref{eq:etaapprox}) is technically valid not only at the end of inflation but merely when $m
_Q\gg1$ and therefore it is approximately also valid at the onset of backreaction from right handed modes even though that occurs much before the end of inflation.}. At least in the cases for which $g\sim{\cal O}(10^{-4})$ which are relevant to our analysis, since one needs at least a value of $m_Q\sim -5$ to make the right-handed mode backreaction relevant. 
This is the reason we do not pursue numerically simulating the right handed mode backreaction in the present work.

We stress that we also checked against the possibility of encountering large $\eta_{\chi}$ values 
suppressing perturbations in the weak regime of CNI, finding that this is not the case. More specifically, in the weak CNI regime one obtains an expression of the form $\eta_{\chi,{\rm END}}\simeq 1/2\cdot {\cal O}(\epsilon_H)(m_Q^2 \Lambda^2)/(1+m_Q^2)$ which, at least parametrically, matches the bound in Eq. (\ref{eq:etalimit}), as it should.

Finally, at this stage that our analysis does not take into account the scalar perturbations of the metric. These perturbations have only a small impact on the analysis of CNI in the weak backreaction regime \cite{Dimastrogiovanni:2012ew,Adshead:2013nka}.
We are operating under the reasonable assumption that this conclusion extends to the case of strong backreaction. Such expectation is supported by the fact that all coupling terms between metric perturbations and axion/gauge fields are slow-roll suppressed. This leads to the conclusion that metric perturbations will not significantly affect the evolution of the modes as long as the slow roll parameters $\epsilon_H$ and $\eta_H$ remain small, which is certainly the case in our work.

%
\section{Pure Chromo-Natural Inflation and a sudden end to inflation}
\label{sec:pure-chromo}
%
In our analysis so far we have striven to be as agnostic as possible about the details of the potential, we only asked that it has a minimum at $V(\chi)=0$. We shall now  consider more concrete potentials and present a simple toy model which enables us to end inflation while circumventing the challenges discussed in the previous section. Since such challenges boil down to inflation lasting too long a time in the strong backreaction regime, the simplest way to overcome them is to assume an inflationary potential with a sudden transition to zero at some specific field value $\chi_{\rm crit}$. Such potentials have already been considered in the axion inflation literature \cite{Hashiba:2021gmn}, are known to be compatible with reheating \cite{Spokoiny:1993kt}, and resemble the potentials of quintessential inflation \cite{Peebles:1998qn}. The potential we consider obeys the following
\begin{equation}
 V(\chi) =
  \begin{cases}
  V_{\rm Inf}(\chi) & \text{for} \;\;\;\chi \leq \chi_{\rm crit} \\
  0 & \text{for} \;\;\;\chi > \chi_{\rm crit}\;.
  \end{cases}
\end{equation}

In such scenario inflation ends instantaneously at $\chi_{\rm END}=\chi_{\rm crit}$ and subsequently, after a very brief period of kination, the universe is dominated by the gauge field perturbations which are continuously being produced during inflation due to the Chern-Simons coupling. Finally, the universe, dominated by dark radiation, simply expands until interactions between the dark radiation and the standard model reheat the universe. Alternatively, if the non-Abelian gauge field is confined, dark glueballs may form which can then decay into the standard model thereby reheating the universe. We leave concrete realizations of reheating in this scenario to future work. 

The field value at which the sudden transition happens is chosen so that a total of 60 e-folds of inflation are achieved while the geometrical $\eta_{\chi}$ parameter is small throughout inflation, and the strong backreaction from the right-handed modes is negligible. 

The last ingredient that is left to specify is the choice of the inflationary potential before the sudden transition. Instead of the conventionally chosen sinusoidal potential, we will instead explore here potentials with a flat plateau. 
In this work we employ the \textit{pure natural inflation} potential \cite{Nomura:2017ehb} defined as
\begin{equation}
    V_{\text{PNI}}(\chi) =M^4\left[ 1 -\left(1 
    + \left(\chi/F\right)^2\right)^{-p}\right]\;.
    \label{eq:PNIP}
\end{equation}
\begin{figure}
\begin{center}
\includegraphics[scale=0.7]{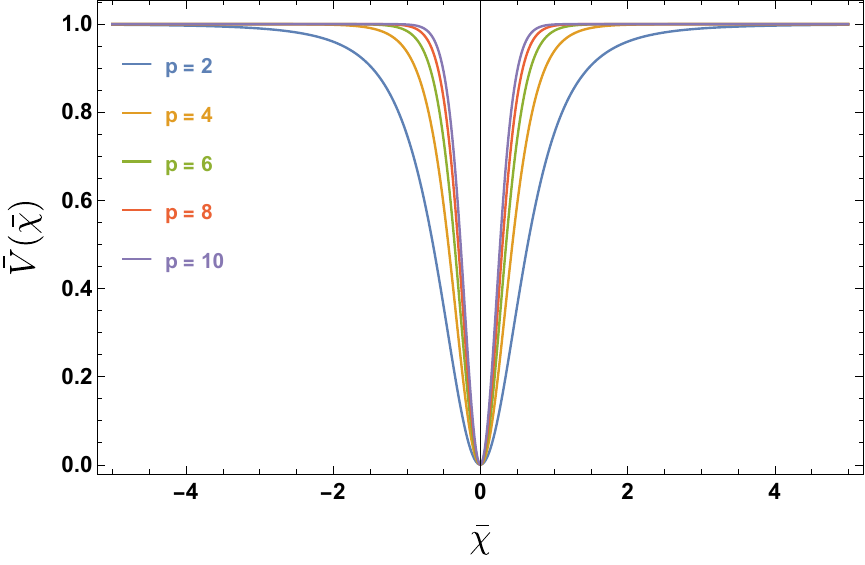}
\end{center}
\vspace*{-5mm}\caption{Plot of the potential in (\ref{eq:PNIP}) for various values of $p$. We have rescaled $\bar{\chi}\equiv\chi/F$ and $\bar{V}\equiv V/M^4$.}
\label{fig:PNIpot}
\end{figure}

This is a typical non-periodic potential that is expected to stem from non-perturbative effects due to the coupling of an ALP to an SU($N_c$) field in the large $N_c$ limit \cite{tHooft:1973alw,Witten:1980sp,Witten:1998uka,Nomura:2017ehb,Nomura:2017zqj}. Note that we do \textit{not} identify this field with the SU(2) field whose coupling to the ALP has been written explicitly in our action eq.~(\ref{eq:CNIL}) which is at weak coupling. Such models feature potentials with a multi-branch structure each of which is characterized by a flat plateau at large field values. We plot the potential in Fig.~\ref{fig:PNIpot} for a single branch for various values of $p$. The parameters $F$ and $M$ have mass units and are related to the axion decay constant and potential mass via the approximated relations \cite{Nomura:2017ehb}
\begin{align}
    M^4 \approx \sqrt{N_c}\mu^4
    &&
    F \approx N_c f 
\end{align}
where $N_c$ is the number of colors. As a consequence of this relation, we can use $F>M_p$ in our numerical implementation while still having a sub-Planckian axion decay constant. The parameter $p$ is dimensionless and, a wide range of its value has been explored in related literature  \cite{Chatrchyan:2023cmz}. For convenience, we  define the effective parameter
\begin{equation}
    \tilde{\lambda}\equiv \lambda/N_c
\end{equation}
so that we can write the Chern-Simons coupling using either one of the following two equivalent expressions 

\begin{equation}
    \frac{\lambda}{4 f}\chi F_{\mu\nu}^a\tilde{F}^{a\, \mu \nu} \equiv \frac{\tilde{\lambda}}{4 F}\chi F_{\mu\nu}^a\tilde{F}^{a\, \mu \nu}\;.
\end{equation}

As alluded to earlier in this section, we pick this type of potential as opposed to a periodic one because the flat plateau at large field values is expected to feature better  in terms of compatibility with CMB observations. The sinusoidal potential  has instead been  ruled out even when accounting for non-linear contributions \cite{Papageorgiou:2018rfx}. However, previous analyses of the periodic potential are incomplete since (i) these only explored the large $\Lambda$ limit whereas more solutions become possible if one considers $\Lambda\simeq {\cal O}(1)$ and given that (ii)  they disregarded the role of strong backreaction in prolonging the duration of inflation. If such effects are taken into full consideration, it turns out to be possible for the periodic potential to become compatible with CMB measurements. The price to pay is in the form of fine-tuned initial conditions, i.e. the axion ought to start its dynamics very close to the maximum of the potential.

Instead of pursuing the cosine scenario, we  choose to explore non-periodic potentials; these do not require any engineered range of values for the axion initial conditions. 
In the next section we dive into the phenomenology of the model and the related signatures.

\section{Phenomenology and Signatures}
\label{sec:pheno}

Having built up all the machinery necessary to study the phenomenology of this model, we  study here on a key case exemplifying the sort of signatures one can expect. Our focus will be on the linear contributions to the scalar and tensor power spectra at small (i.e. interferometer) scales. 

To begin with, we demand the scalar power spectrum is compatible with CMB observations, thus considerably reducing the allowed parameter space. We describe in full a useful  prescription to implement these bounds in Appendix~\ref{app:density_perturbation}. Assuming the initial dynamics sits in the weak backreaction regime, the model under consideration has a set of six free parameters, namely $(\lambda,g,\mu^4,F,p,\chi_{\rm CMB})$ where $\chi_{\rm CMB}$ is the axion  value when CMB scales cross the horizon. It turns out to be conveninent to enforce the CMB constraints on the scalar power spectrum by acting on the parameters $(g,m_{Q,{\rm CMB}},\Lambda_{\rm CMB})$, which in turn fixes three of the six  parameters listed above\footnote{Selecting which parameters to act on in order to secure agreement/compliance with a given measurement/constraint is of course generically a matter of choice. Nevertheless, the parameter space one arrives at upon satisfying/implementing all existing measurements/constraints is unique so long as one employs a consistent algorithm.}. In our algorithm, the three remaining parameters after enforcing CMB constraints are $\lambda$, $p$ and $\mu^4$. The selection of the range of interest for the three residual parameters is made with two criteria in mind.\\
(i) We require that the transition from weak to strong backreaction regime starts out no later than about $N_{\rm trans}\sim 35$ e-folds before the end of inflation. Most of the particle production takes place during the transition phase and therefore if the transition takes place in the $25-35$ e-folds interval before the end of inflation  the signatures corresponding to particle production will be within reach of next generation interferometers.\\
(ii) We ask that throughout 60 e-folds of evolution the geometrical parameter $\eta_\chi$ remains safely within the bounds computed in Sec.~\ref{sec:largeeta}. 

The first requirement restricts the acceptable range of parameters quite considerably. This is easy to see since, to require that strong backreaction becomes relevant $\sim 25$ e-folds after the CMB mode left the horizon implies a rapid change in the evolution of the model variables: from nearly constant as necessary at CMB scales, to rapidly increasing just 25 e-folds later. 

When systematically studying the parameter space, we have used the well-known fact (both analytical and lattice evidence exists) that a strong backreaction regime delays the end of inflation with respect to the estimate extrapolated from an initial weak backreaction configuration. Anticipating such delay due to strong backreaction, we demanded that the CMB scale crossed the horizon (in the weak backreaction regime and) much less than the standard $60$ e-folds before the end of inflation. One quickly gets an intuition and is able to estimate the delay due to the intervening strong backreaction so that the full evolution, including both regimes, delivers the desired 60 e-folds.

\begin{figure}
\begin{center}
\includegraphics[scale=0.7]{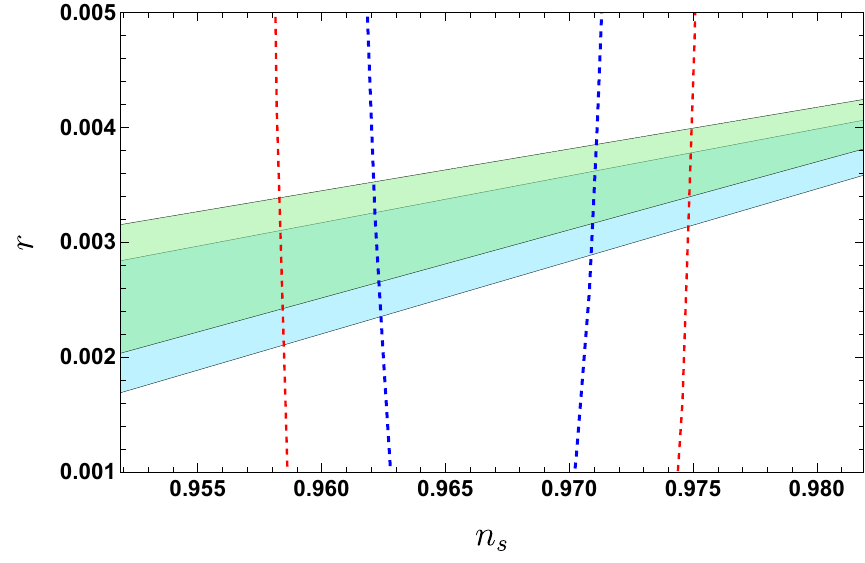}
\end{center}
\vspace*{-5mm}\caption{Expected values of $r$ and $n_s$ in the PCNI scenario. The shaded areas are for $p= 4,10$ (blue and green respectively). The top and bottom lines, for each region, represent approximately $30$ and $45$ e-folds of evolution in the weak backreaction attractor, while the left-to-right variation corresponds to different values for $m_{Q,{\rm CMB}}$ (from $2.35$ to $2.5$). The dashed lines designate the 1 and 2 $\sigma$ contours of the joint Planck-Bicep analysis \cite{BICEP:2021xfz}. }
\label{fig:rnsPlane}
\end{figure}

The result of this analysis for two indicative choices of the potential parameter $p$ is shown in Fig~\ref{fig:rnsPlane}. For the purpose of generating this plot, we fixed $\Lambda_{\rm CMB}=2$. In all the cases studied the strong backreaction regime delays the end of inflation ensuring that a $\chi_{\rm crit}$ can be found with a total duration of inflation of $60$ e-folds. We arrive then at the following set of fiducial parameters

\begin{align}
&\tilde{\lambda}=760\;, & &g=6.5\times10^{-4}\;, & &M^4=4.9\times 10^{-11}\;M_p^4\;, \nonumber\\
&F=7.2 \;M_{p}\;, & &p=9\;, & &\chi_{\rm CMB}=-3.95 \;M_p\;.
\label{eq:fid}
\end{align}

\begin{figure}
\begin{center}
\includegraphics[scale=0.5]{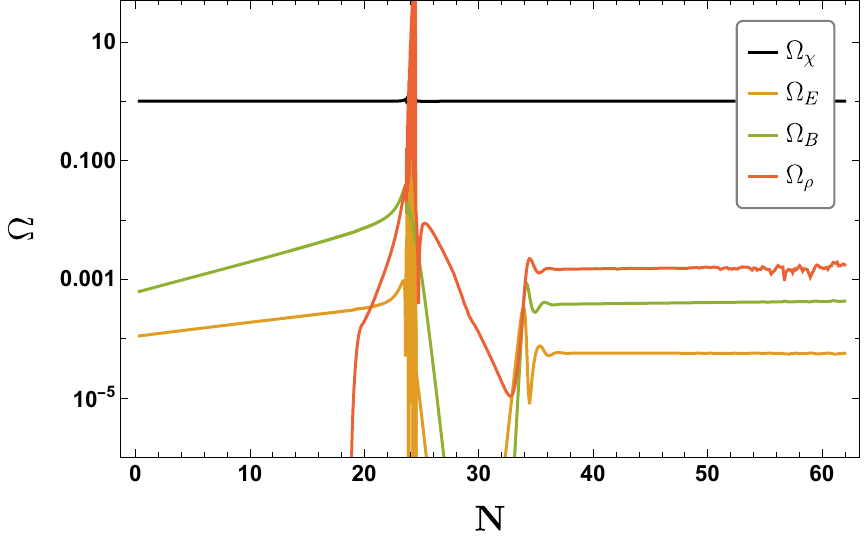}
\includegraphics[scale=0.5]{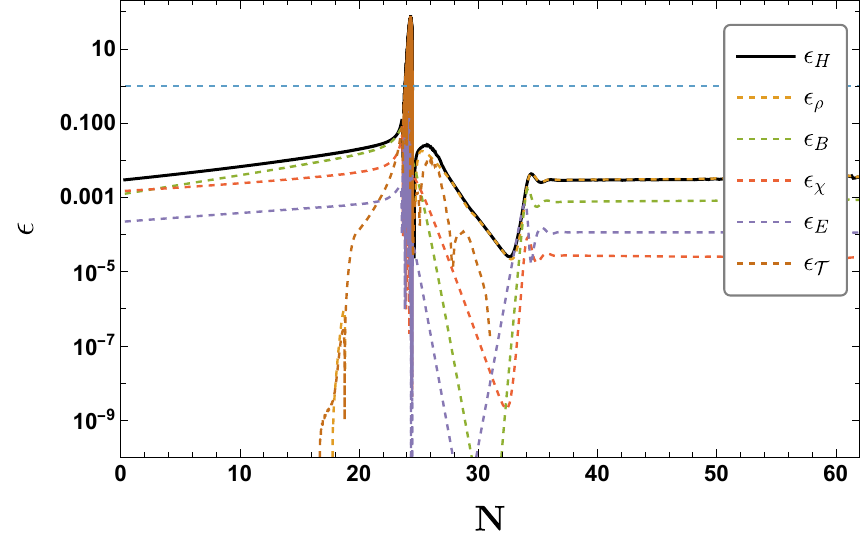}
\end{center}
\vspace*{-5mm}\caption{The left panel displays the fractional energy densities as functions of time. It is clear that the inflaton makes up the dominant component in the energy density of the universe at all times except for a brief period during the transition. On the right panel the various slow-roll parameters are being displayed with a black solid line representing the sum of the individual contributions as defined in eq.~(\ref{eq:slow}).}
\label{fig:epsilonomega}
\end{figure}

Note that even though the potential scale $F$ is trans-Planckian, the axion decay constant can be sub-Planckian for large enough number of colors $N_c\gtrsim7$. These parameters  allow for the transition to take place sufficiently early and also ensure that the $\eta_\chi$ parameter is small enough until the potential runaway ends inflation 60 e-folds after the CMB modes crossed the horizon. We evolve the background numerically along with the tensor perturbations and include the backreaction at the homogeneous level. Our numerical approach parallels that used in \cite{Cheng:2015oqa,Notari:2016npn,DallAgata:2019yrr,Garcia-Bellido:2023ser} (see also: \cite{Sobol:2019xls,Domcke:2020zez,Gorbar:2021rlt,vonEckardstein:2023gwk,Domcke:2023tnn} for other approaches). The evolution of the particle production parameters $m_Q$ and $\xi$ for these fiducial parameters are shown in Fig.~\ref{fig:transition}. Additionally, we display the evolution of the fractional energy densities and the various slow roll parameters for the same fiducial example in Fig.~\ref{fig:epsilonomega}.

\begin{figure}
\begin{center}
\includegraphics[scale=0.7]{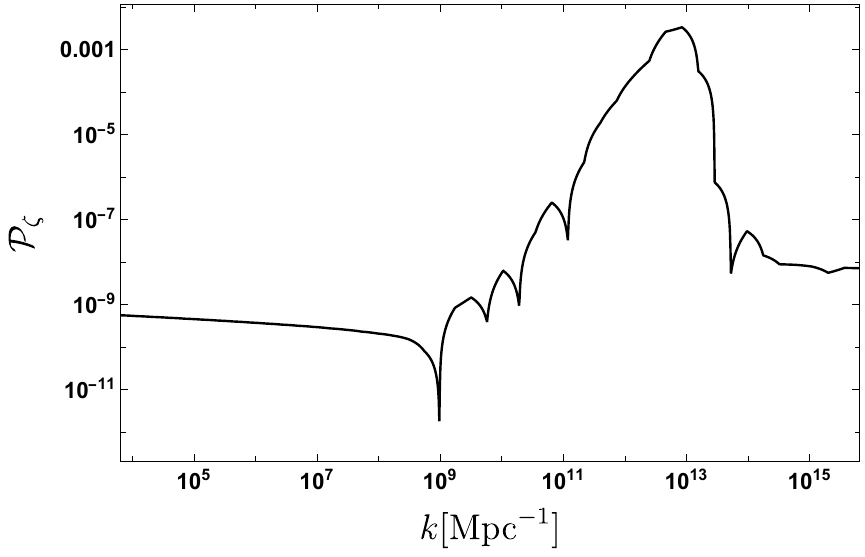}
\end{center}
\vspace*{-5mm}\caption{Power spectrum of the density perturbation for the fiducial parameters listed in (\ref{eq:fid}). The modes of production of this peaked spectrum are explained in the last paragraph of Sec.~\ref{sec:transition}.}
\label{fig:Pz}
\end{figure}

As emphasized in Sec.~\ref{sec:transition}, the transition from weak to strong backreaction regime engages two mechanisms that result in powerful production of scalar perturbations ensuring that a peaked spectrum is generated. The location of the peak tracks   the moment when the transition occurs during inflation. The location is therefore very much dependent on the parameters that shape the weak backreaction regime. The peak amplitude may be directly enhanced (or suppressed) by changing $F$ and $p$.  It is in principle possible to support a power spectrum amplitude close to $10^{-2}$, which is the scale required for a level of PBH production that make up the entirety of dark matter (in the monochromatic case and assuming Gaussian sourcing \cite{Dalianis:2018frf}). Using the formalism in Appendix~\ref{app:scalar} we compute the power spectrum for the fiducial example and plot it in Fig~\ref{fig:Pz}. The  spectrum is peaked at scales of about $k\sim 10^{13} \;{\rm Mpc}^{-1}$ with an amplitude of order $\mathcal{P}_\zeta\sim 10^{-3}$. 

Let us now move on to the GW signal in the same scenario. We expect at least three contributions to the final spectrum on top of the inevitable vacuum one. The first one is the scalar-induced contribution associated to the scalar perturbations discussed in the previous paragraph. We computed the scalar-induced gravitational wave signal numerically, using SIGWfast \cite{Witkowski:2022mtg}. Additionally, at the end of phase I the large magnitude of the particle production parameters $m_Q$ and $\xi$ imply a strong direct (i.e. linear) GW sourcing from  tensor perturbations in the gauge sector, see Eqs.~(\ref{eq:tmodes}) and~(\ref{eq:tmodes}). Finally, during phase IV, the same tachyonic instability produces sourced GWs of the opposite polarization but with a much smaller amplitude. The total gravitational wave signal evaluated today can be written as \cite{Caprini_2018}
\begin{equation}
    \Omega_{\text{GW,0}}(k) = \frac{3}{128}\Omega_{\text{rad},0} P_{h}(k)\left[\frac{1}{2} \left(\frac{k_{\rm eq}}{k}  \right)^2 +\frac{16}{9}  \right]\;,
\end{equation}
where $P_h$ is the sum of the individual contributions, $\Omega_{\rm rad,0}=h^{-2}2.47\times 10^{-5}$, $\Omega_{\rm m,0}=h^{-2}0.14$, $h=0.674$ and $k_{\rm eq}=2\pi a_0 H_0 \Omega_{\rm m,0}/(\pi\sqrt{2\Omega_{\rm rad,0}})$.

\begin{figure}
\begin{center}
\includegraphics[scale=0.7]{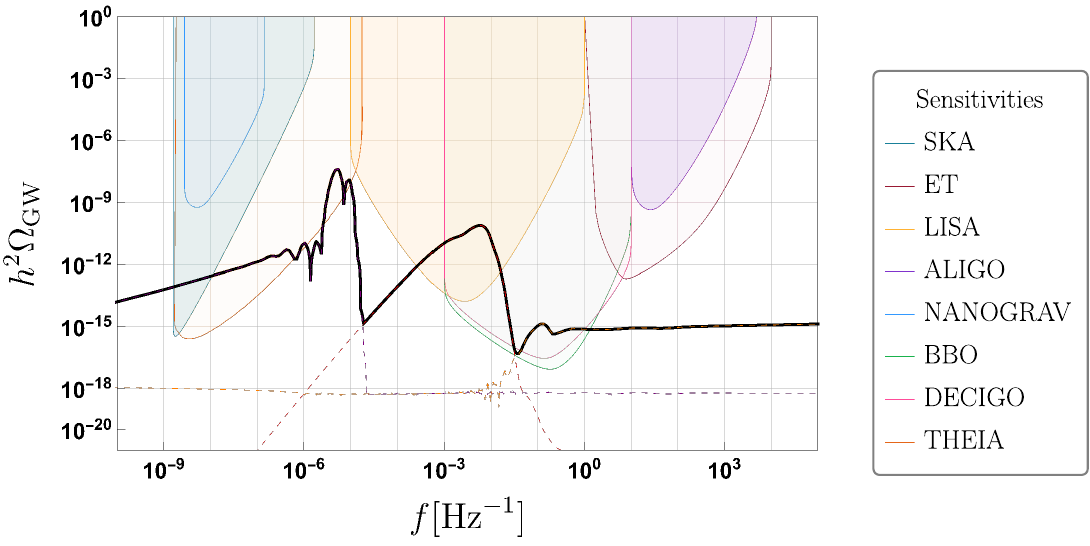}
\end{center}
\vspace*{-5mm}\caption{Total GW signal for the fiducial parameters listed in (\ref{eq:fid}). The black solid line represents the sum of all contributions. The purple and orange dashed lines are the contributions of the left and right tensor modes, respectively. The red dashed line is the contribution of the scalar induced part of the spectrum.}
\label{fig:gW}
\end{figure}

The final result of the full spectrum as well as  its individual contributions is shown in Fig.~\ref{fig:gW}. There are several noteworthy features in this plot. First, the full spectrum features a series of well-defined peaks. The first peak (around $f\sim 10^{-5}$) occurs as a consequence of the direct sourcing of gravitational waves from the unstable $t_L$ mode produced during phase I. The second peak (around $f\sim 10^{-2}$) is the scalar-induced contribution. Finally, the small peak (around $f\sim 10^{-1}$) shows the onset of the strong backreaction attractor and the onset of instability of the right-handed mode $t_R$. As is clear from Fig.~\ref{fig:gW}, our fiducial set of parameters supports a GW signal detectable by LISA as well as more futuristic experiments.

The distance between the three peaks stays roughly constant as one scans the parameter space but the position itself of the three peaks may change significantly depending on the set of parameters chosen. Another noteworthy feature of the GW spectrum is that the part that is indicated by the purple line is totally chiral and left-handed, the red one is unpolarized, and the orange part is purely right-handed. This feature is a smoking gun for the transition from weak to strong backreaction in this and similar CNI-like scenarios. The measurement of the net circular polarization (and thus chirality) of a stochastic GW background by future experiments is indeed a realistic \cite{Domcke:2019zls,Mentasti:2023gmg} prospect as well as a very intriguing one in light of our findings here. 

\section{Conclusions}
\label{sec:conclusion}

In recent years, there has been remarkable progress in understanding the dynamics between axions and gauge fields during inflation. Despite these successes, several open questions remain, ranging from model-building and higher-dimensional embeddings of the inflationary mechanism to strong backreaction dynamics and its heightened relevance during the later stages of inflation.

In the present work, we considered a model of axion inflation coupled to a non-Abelian gauge sector featuring a homogeneous and isotropic background. We go beyond previous studies in several respects. First, we studied non-periodic potentials in the presence of the gauge field background which arise in the large $N_c$ limit of SU($N_c$) theory coupled to axions at strong coupling \cite{Nomura:2017ehb}. This setup, which we refer to as pure chromo-natural inflation, fares better vis-\'{a}-vis CMB observations than standard CNI as it features a flat plateau at large field values. Furthermore, building on previous work \cite{Dimastrogiovanni:2024xvc}, we explored in detail the various particle production mechanisms in this setup 
and numerically studied the field evolution during inflation, starting from an early period characterized by weak backreaction from field fluctuations. 

We find that these models generically transition from weak to strong backreaction at some point before the end of inflation. Our analysis underscores the importance of accounting for the extended duration of inflation due to strong backreaction, as this is crucial for a consistent implementation of CMB constraints. The transition from one regime to another typically leaves behind strong imprints in the form of GW production, which may be observable by next-generation interferometers such as LISA, as well as scalar perturbations that may yield significant PBH production. 
In the case of GWs, we find that they exhibit unique chiral characteristics and a rich peak structure whose spacing can be highly insensitive to the underlying parameter space.

Besides seeking further, concrete, mechanisms to end inflation alternative to the option we present in Section \ref{sec:pure-chromo}, one might wonder about the post-inflationary evolution of our system, starting with reheating. Indeed, the in-depth study of such dynamics is left to future work as details will depend crucially on how the non-Abelian gauge sector couples to the Standard Model.  Another interesting aspect we should comment on here is the inclusion of recently derived  perturbativity constraints \cite{Dimastrogiovanni:2024lzj} (see also \cite{Ferreira:2015omg,Peloso:2016gqs}).  
Depending on the parameter space in hand, such constraints can be competitive with backreaction limits. It follows that only a full numerical analysis of both effects, preferably on the lattice, can ascertain whether strong backreaction is always accessible without running into perturbativity bounds first.

Finally, although scalar backreaction is expected to be sub-leading for the field content of our model, it would be very interesting to study its impact on the evolution of the background equations. Notably, in the presence of spectator fields \cite{Dimastrogiovanni:2024xvc}, scalar backreaction can significantly contribute to the phenomenology of axion–gauge field models. Given that such an analysis is expected to be much more resource-intensive than the present work, we leave a combined study of tensor and scalar backreaction to future investigations.

In closing, we would like to highlight another promising direction for future research. Throughout this work, we have developed a numerical framework tailored for the study of the strong backreaction regime. It is sufficiently agile to find application both in the Abelian and non-Abelian cases. A natural next step would be to apply our analysis to the case of multiple spectator axion sectors, as found in the so-called axiverse \cite{Arvanitaki:2009fg,Acharya:2010zx,Cicoli:2012sz,Demirtas:2018akl,Demirtas:2021gsq, DAmico:2021vka,DAmico:2021fhz,Dimastrogiovanni:2023juq}.
 Existing work on the gravitational wave “forest” \cite{Kitajima:2018zco} in such scenarios are typically \cite{Dimastrogiovanni:2023juq} confined to the weak backreaction regime, whereas our framework  enables exploration well beyond this limit.

\subsection*{Acknowledgments}
We are delighted to thank Mattia Cielo  for very helpful discussions.~MF~and~AP~acknowledge the “Consolidaci\'{o}n Investigadora” grant CNS2022-135590.~The work of MF, AP, and CZ is partially
supported by the Spanish Research Agency (Agencia Estatal de Investigaci\'{o}n) through the Grant IFT Centro de Excelencia Severo Ochoa No CEX2020-001007-S,
funded by MCIN/AEI/10.13039/501100011033. MF acknowledges support from the “Ram\'{o}n y Cajal” grant
RYC2021-033786-I.

\appendix

\section{Scalar perturbations equations of motion}
\label{app:scalar}

The equations of motion for the scalar perturbations in CNI have been already derived in several works~\cite{Dimastrogiovanni:2012ew,Adshead:2013qp,Adshead:2013nka,Papageorgiou:2018rfx,Dimastrogiovanni:2024xvc}. However, some of the results presented above either assume an exact DeSitter approximation  or assume the weak backreaction regime of CNI for which $\xi=m_Q+1/m_Q$. For completeness we write the full equations without an exact DeSitter or weak backreaction regime assumption. This allows for our equations to be used even if the universe stops inflating for some short period and also in the strong backreaction regime of CNI. It has been shown in several works that integrating out the scalar perturbations of the metric provide negligible contributions to the equations of motion of the dynamical scalar perturbations in CNI. This is not too surprising as metric perturbations couple only gravitationally while the scalar perturbations of the SU(2) multiplet and the inflaton perturbation couple directly. For this reason we will disregard scalar metric perturbations ($\delta g_{ij}|_{\rm scalar}=0$) and proceed to study only the SU(2)-inflaton perturbation system. The scalar perturbations of the SU(2) multiplet are arranged as follows

\begin{align}
    \delta A^1_\mu&=[0, \delta \phi(t,z)-Z(t,z), \chi_3(t,z),0]\;,\nonumber\\
    \delta A^2_\mu&=[0,-\chi_3(t,z),\delta\phi(t,z)-Z(t,z),0]\;,\nonumber\\
    \delta A^3_\mu&=[\delta A^3_0(t,z),0,0,\delta\phi(t,z)+2Z(t,z)]\;.
\end{align}

The perturbation of the inflaton is simply $\delta\chi(t,z)$. We have chosen to align the momentum of the perturbations with the $z$-axis. We are free to do that without loss of generality at the linearized level. 

Gauge fixing allows us to remove one of the perturbations listed above. We fix the gauge such that

\begin{equation}
    \chi_3=i k \frac{2Z+\delta\phi}{2 g a Q}\;.
\end{equation}

Once expanding the Lagrangian to the quadratic level in perturbations it is possible to see that the mode $\delta A^3_0$ is non-dynamical and can be directly integrated out with the constraint equation

\begin{equation}
\delta A^3_0=-\frac{4g a^2(H Q+ \dot{Q})\chi_3+g a^2 Q^2 \frac{\lambda}{f}(-ik)\delta\chi}{k^2+2g^2 a^2 Q^2}\;.
\end{equation}

The residual Lagrangian can be written in terms of perturbations $\delta\chi, \delta \phi $ and $Z$. we perform one final transformation to make the kinetic terms canonical 

\begin{equation}
    \hat{X}\equiv a \delta\chi\;\;\;,\;\;\;\hat{Z} \equiv \sqrt{2}(Z-\delta\phi)\;\;\;,\;\;\; \hat{\varphi} \equiv \sqrt{2+\frac{k^2}{g^2 a^2 Q^2}}\left(\frac{\delta\phi}{\sqrt{2}}+\sqrt{2}Z\right)
\end{equation}

We opt to display the equations with respect to physical time for concreteness and we are omitting the arguments of the time-dependent variables.

\newcommand{\XX}{\hat{X}}
\newcommand{\PS}{\hat{\varphi}}
\newcommand{\ZZ}{\hat{Z}}
\newcommand{\dem}{k^2 + 2  m_Q^2\,a^2 H^2 }
\begin{align}
    \ddot{\XX} & + H \dot{\XX} + 
    \left[
    -2 + \epsilon_H + 3 \eta_\chi + 
    \frac{ k^2\left(k^2 +  m_Q^2 (2 + \Lambda)\,a^2 H^2 \right)}{a^2 H^2\left(k^2 + 2  m_Q^2\,a^2 H^2\right)} \right] H^2\XX+\frac{\sqrt{2} m_Q^2 \Lambda\, aH}{\sqrt{k^2 + 2  m_Q^2\,a^2 H^2}} H\dot{\PS} \nonumber\\
    &+    
   \frac{m_Q\Lambda\left(k^4 + 3m_Q^2\, k^2  a^2H^2 + 4 m_Q^4\,  a^4H^4  \right) }{aH \left(k^2 + 2 m_Q^2\,a^2 H^2\right)^{3/2}} \sqrt{\frac{2\epsilon_E}{\epsilon_B}}H^2\PS - \sqrt{2} m_Q \Lambda\, H \dot{\ZZ} -
   2 m_Q^2\Lambda  \sqrt{\frac{2\epsilon_E}{\epsilon_B}}\,H^2 \ZZ  = 0,
   \label{eq:scalarpertX}
\end{align}
\begin{align}
    \ddot{\PS} &+ H \dot{\PS} +  
    \left[ 
    \frac{6 m_Q^4\,k^2\, a^2 H^2}{ \left(\dem\right)^2}\frac{\epsilon_E}{\epsilon_B} +
    \frac{k^4 + 2 m_Q(3m_Q - \xi)k^2\, a^2H^2 + 4m_Q^4\, a^4 H^4}{\left(\dem\right)\,a^2H^2}
    \right]H^2 \PS \nonumber\\
    &- \frac{2 (m_Q - \xi)\sqrt{\dem}}{a H}\, H^2\ZZ - \frac{\sqrt{2} m_Q^2 \Lambda\, a H}{\sqrt{\dem}}H\dot{\XX} 
    \nonumber\\
    &
    +\frac{\sqrt{2} m^2_Q \Lambda\, a H}{\sqrt{\dem}}\left[1+
    \frac{ k^4}{m_Q\left(\dem\right)\, a^2 H^2}\sqrt{\frac{\epsilon_E}{\epsilon_B}} 
    \right] H^2 \XX = 0,
    \label{eq:scalarpertPsi}
\end{align}
\begin{align}
    \ddot{\ZZ} &+ H \dot{ \ZZ} +  \left[\frac{k^2}{a^2H^2} + 2m_Q \left(m_Q - 2 \xi\right)   \right]H^2 \ZZ 
    + \sqrt{2} m_Q \Lambda\, H \dot{\XX} - \sqrt{2} m_Q \Lambda\, H^2 \XX
    \nonumber \\
    & -
    \frac{2 \left(m_Q - \xi\right)\sqrt{\dem} }{aH}H^2\PS = 0.
    \label{eq:scalarpertZ}
\end{align}

Note that in the limit $\epsilon_E\rightarrow \epsilon_B/m_Q^2$ and $\epsilon_H\rightarrow 0$, $\eta_\chi \rightarrow 0$ one recovers exactly the system of equations in App. A of~\cite{Dimastrogiovanni:2024xvc}.

The initial conditions for the set of equations above have been analyzed in detail in~\cite{Adshead:2013nka,Papageorgiou:2018rfx} using the WKB approximation for an approximate solution deep inside the horizon. We simply quote the result of this analysis here.

\begin{align}
    \hat{X}_{\rm in}&=\frac{\sqrt{1+m_Q^2}}{\sqrt{2k}}{\rm e}^{i\left(x-x_{\rm in}\right)}\left(\frac{x_{\rm in}}{x}\right)^{i\sqrt{\frac{1+m_Q^2}{2}}\Lambda}\nonumber\\
    \hat{\varphi}_{\rm in}&=-\frac{1}{\sqrt{2k}}{\rm e}^{i\left(x-x_{\rm in}\right)}\left(\frac{x_{\rm in}}{x}\right)^{i\sqrt{\frac{1+m_Q^2}{2}}\Lambda}\nonumber\\
    \hat{Z}_{\rm in}&=\frac{i m_Q}{\sqrt{2k}}{\rm e}^{i\left(x-x_{\rm in}\right)}\left(\frac{x_{\rm in}}{x}\right)^{i\sqrt{\frac{1+m_Q^2}{2}}\Lambda}\;.
\end{align}

The WKB analysis that results in the initial conditions above holds regardless of whether one is analyzing the weak or strong backreaction regime. Instead the terms that dominate deep inside the horizon are entirely independent of $\xi$ and therefore provided we set the initial conditions deep inside the horizon, we can use the same initial conditions. However, it is important to note that for some of the examples we analyze in the main text, parameter $\Lambda$ might be less than order one for some period of time. In such cases the initial conditions can be called into question, however this will never affect the scales for which particle production is strong and hence the relevant phenomenology valid. 

\subsection{Instability in the scalar sector}

It was originally shown in~\cite{Dimastrogiovanni:2012ew} that there is an instability in the scalar sector for $|m_Q|<\sqrt{2}$. By a rigorous application of the WKB approximation in~\cite{Adshead:2013nka} it was demonstrated that the system evolves as a superposition of a slow and fast frequency. Assuming a solution of the type $\sim {\rm e}^{i S(x)}$ for the perturbations deep inside the horizon, the exponent obeys

\begin{equation}
   S'(x)\simeq 1+\frac{3 m_Q^2 \Lambda^2}{2x^2}\mp\frac{\Lambda\sqrt{8(1+mQ^2)x^2+9m_Q^4 \Lambda^2}}{2 x^2}
\end{equation}

Unlike the fast mode, the frequency squared of the slow mode may become zero and even flip sign for a finite value of x. This signals the onset of the instability. Setting the slow mode frequency to zero and solving for x we find.

\begin{equation}
    x_{\rm inst}=\sqrt{2-m_Q^2}\Lambda
\end{equation}

which demonstrates both the conditions for the formation of the instability ($|m_Q|<\sqrt{2}$) as well as the exact amount inside the horizon the mode has to be to start experiencing the instability. Subsequently we can perform a Taylor expansion for parameter $x\ll x_{\rm inst}$ to find that the slow mode frequency is exactly constant (to zeroth order in slow-roll) and equal to 

\begin{equation}
    S'(x)=\Omega_{\rm slow}=\pm \sqrt{\frac{m_Q^2-2}{3m_Q^2}}
\end{equation}

Since the derivative of $S(x)$ is constant, the total value of the exponent can be integrated from the onset of the instability until some late time to give an overall boost to the perturbation amplitude of

\begin{equation}
    {\rm e}^{i S(x_{\rm late})}={\rm e}^{i \int_{x_{\rm inst}}^{x_{\rm late}}\Omega_{\rm slow}}={\rm e}^{\sqrt{\frac{2-m_Q^2}{3m_Q^2}}\sqrt{2-m_Q^2}\Lambda}\label{eq:instability}
\end{equation}

The formula above suggests that one can keep the instability under control, and perhaps even harness it for the purpose of phenomenology, by selecting parameters that don't make the amplitude too high. The typical parameters that are being used in CNI involve values of $m_Q$ that stay of ${\cal O}(1)$ in order to not overproduce tensor perturbations. However, parameter $\Lambda$ has been shown less attention in the literature. For the most part past analyses have chosen parameters such that $\Lambda \sim {\cal O}(10-100)$ and therefore one can have successful CNI regime. Such large values of $\Lambda$ are incompatible with the instability analysed here as the exponent of~(\ref{eq:instability}) explodes in a way that is not under control. However, CNI can still be realized for more modest values of $\Lambda\sim 2-3$ since they still satisfy the CNI condition $\Lambda>\sqrt{2}$. In this case, and additionally if the $|m_Q|<\sqrt{2}$ traversal takes place for a limited number of e-folds, the instability can be under control and even useful for producing novel signatures in the scalar sector.

\section{Density perturbation in terms of scalar field fluctuations and value of $P_\zeta$}
\label{app:density_perturbation}

We dedicate this Appendix to deriving anew the relationship between the gauge invariant density perturbation and the scalar perturbations in our model. In the flat gauge we can write
\begin{equation}
\zeta=\frac{1}{3(\rho+P)}\left(\delta T^0_{\;\;0}\right)^{\rm flat}\;.
\end{equation}
As emphasized in the previous Appendix, we are computing the density perturbations in the absence of the scalar perturbations of the metric. Under this assumption it is straightforward to proceed using the quadratic action. In terms of physical fluctuations we get

\begin{align}
    \zeta
    =\frac{1}{6 \epsilon_H \,H^2 M_p^2}
    \Bigg[
    V'(\chi)\delta\chi +
    \dot{\chi}\dot{\delta\chi}+
    3\big(H^2 Q &+2 g^2 Q^3 +H \dot{Q}\big)\left(\frac{\delta\phi}{a}\right) \nonumber \\ 
    &+3\left(H Q+\dot{Q}\right)\frac{d}{dt}\left(\frac{\delta\phi}{a}\right)+{\cal O}\left(\frac{k^2}{a^2}\right)
    \Bigg]\;,
\end{align}
or in terms of canonically normalized fluctuations we obtain
\begin{align}
    \zeta=\frac{1}{6 \epsilon_H \,H^2 M_p^2}\Big[&V'(\chi)\left(\frac{\hat{X}}{a}\right)+\dot{\chi}\frac{d}{dt}\left(\frac{\hat{X}}{a}\right)+\left(H^2 Q+2 g^2 Q^3+H \dot{Q}\right)\left(\frac{\hat{\varphi}}{a}\right)+(H Q+\dot{Q})\frac{d}{dt}\left(\frac{\hat{\varphi}}{a}\right)\nonumber\\
    &-\sqrt{2}(H^2 Q+2 g^2 Q^3+H\dot{Q})\left(\frac{\hat{Z}}{a}\right)-\sqrt{2}\left(H Q+\dot{Q}\right)\frac{d}{dt}\left(\frac{\hat{Z}}{a}\right)+{\cal O}\left(\frac{k^2}{a^2}\right)\Big]\;.
\end{align}
This is the final formula that is being used to compute the power spectrum as shown in the phenomenology section of the main body of the paper (Sec.~\ref{sec:pheno}). In practice, we are only keeping the terms proportional to $\hat{X}$ as these are far greater than the terms proportional to $\hat{\varphi}$ in the weak backreaction and transition phases respectively.

\medskip

\subsection{Phenomenology of the CMB mode} 

The process of obtaining the measured value for scalar fluctuations of the CMB  deserves special attention. The overall power in scalar perturbations receives at least two contributions. One, derived above, arises at the linear level. A further non-linear contribution exists, it was originally derived in \cite{Papageorgiou:2018rfx}. The total result can be written as follows
\begin{equation}
    P_{\zeta,{\rm CMB}}=\frac{k^3 V'(\chi)^2}{72 \pi^2 H^4 M_p^4 \epsilon_H^2}\frac{|\hat{X}|^2}{a^2}\left(1+R_{\delta\chi}\right) \;\;\;{\rm where}\;\;\;R_{\delta\chi}=6.6\times 10^{-16}m_Q^{18}\,{\rm e}^{2\pi m_Q}\; ,
\end{equation}
where we have kept only the dominant term (the one proportional to $\hat{X}$) in the weak backreaction regime. The formula above can be significantly simplified by assuming an exact DeSitter regime at CMB scales and also replacing $\epsilon_H\simeq \epsilon_B$ as is valid in the weak backreaction regime. The simplified version takes the form
\begin{equation}
    P_{\zeta,{\rm CMB}}=\frac{g^2\Lambda^2}{16 \pi^2 m_Q^4}|x \sqrt{2k}\hat{X}|^2\left(1+R_{\delta\chi}\right)\;.
    \label{eq:Pzetanorm}
\end{equation}
The formula above is what we use to normalize the power spectrum to the observed value $P_{\zeta,{\rm obs}}=2.1\times10^{-9}$.

It is important to note that the quantity $|x \sqrt{2k}\hat{X}|$ becomes nearly constant outside the horizon, as expected. In practice, there is always a small residual running of the superhorizon perturbation which is due to our neglecting of the metric perturbations, as pointed out in \cite{Dimastrogiovanni:2023oid}. In the same reference, it was also indicated  that we can always get a very reliable result for the final, frozen, superhorizon perturbation if we evaluate the mode about two e-folds after horizon crossing. This will be the case as long as the slow roll parameters are much smaller than unity, which certainly to the CMB mode in the context of this work. We will adopt the same approach in the present work. 

Moreover, under the assumption that at very early (CMB scale) times all parameters should be nearly constant, we will be evaluating the CMB mode two e-foldings after horizon crossing by solving the system (\ref{eq:scalarpertX}, \ref{eq:scalarpertZ}, \ref{eq:scalarpertPsi}) assuming constant parameters $m_Q,\;\xi$ and $\Lambda$. In addition, the slow-roll parameters are set to zero. This is a good approximation for both the calculation of $P_\zeta$ and $n_s$ because the variation of the model parameters at such early times is very small compared to the time it takes for a single mode to cross the horizon.

Under the above assumptions we can demand that $P_{\zeta,{\rm CMB}}\simeq P_{\zeta, {\rm obs}}$ and solve for $g$.
\begin{equation}
    g\simeq\sqrt{\frac{16 \pi^2 m_Q^4P_{\zeta,{\rm obs}}}{\Lambda^2|x \sqrt{2k}\hat{X}|^2\left(1+R_{\delta\chi}\right)}}
\end{equation}
The RHS of the above formula depends implicitly, exclusively on $m_Q$ and $\Lambda$. This is an important notion because it reveals that, upon selecting the values $m_{Q,{\rm CMB}}$ and $\Lambda_{\rm CMB}$, the correct value of $g$ that gives the appropriate power spectrum value is unique. We can take this conclusion a step further and consider bounds on the parameters $m_{Q,{\rm CMB}}$ and $\Lambda_{\rm CMB}$ which can give us clues about the acceptable values of the gauge coupling constant $g$. In the case of $m_{Q,{\rm CMB}}$ the inequalities one obtains are as follows
\begin{equation}
    \sqrt{2}<m_{Q,{\rm CMB}}\lesssim 2.6\; ,
    \label{eq:mQlim}
\end{equation}
where the lower limit is to avoid the instability in the scalar modes and the upper limit is required to avoid the nonlinear contribution being greater than the linear one. The second requirement is merely a conservative limit that ensures that the scalar non-Gaussianity of the model remains  within the current constraints \cite{Papageorgiou:2018rfx,Papageorgiou:2019ecb}. In the case of parameter $\Lambda_{\rm CMB}$ the lower limit is well known to be $\sqrt{2}<\Lambda_{\rm CMB}$ so that the gauge field background effective potential has a local minimum distinct from zero. The upper limit is less transparent.

One may set an upper limit in anticipation of transitioning from weak to strong backreaction regime. As we have seen in Sec.~\ref{sec:transition}, such a transition will force the parameter $m_Q$ to become less than $|\sqrt{2}|$, which will inevitably trigger the scalar instability described in \ref{app:scalar}. Since we want scalar perturbations to remain under control, and knowing that their enhanced amplitude is exponentially sensitive to $\Lambda$, we have no choice but to demand that already at CMB scales  $\Lambda_{\rm CMB}$ is no greater than ${\cal O}(1)$. This leaves only a relatively narrow allowed window for $\Lambda_{\rm CMB}$, that is 
\begin{equation}
    \sqrt{2}<\Lambda_{{\rm CMB}}\lesssim {\cal O}(1)\;.
    \label{eq:lambdalim}
\end{equation}
Finally, converting the bounds (\ref{eq:mQlim}) and (\ref{eq:lambdalim}) into a rough estimation on $g$, we find that the  order of magnitude that yields an overall acceptable $P_{\zeta,{\rm CMB}}$ to be 
\begin{equation}
    g\sim {\cal O}(10^{-4})\;.
\end{equation}

To close out this section, our prescription for computing the spectral index $n_s$ consists of finding $P_\zeta$ for a mode that crosses the horizon $60$ e-folds and $60-\delta N$ e-folds before the end of inflation and then numerically computing 

\begin{equation}
    n_s=1+\frac{d\ln P_\zeta}{d\ln k}\simeq 1-\frac{d\ln P_{\zeta}}{dN} \simeq 1-\frac{1}{P_{60}}\frac{P_{60}-P_{60-\delta N}}{\delta N}\; ,
\end{equation}
where $\delta N\simeq 0.1$ was chosen to be sufficiently small so that the result is convergent. The result above disregards the variation of H with respect to that of a.
\section{The $\eta$ parameter in the strong backreaction regime}
\label{app:analytics}

Using the equations of motion for the axion and gauge fields and employing the SB attractor regime condition in Eq.~(\ref{NewAttEq}), one arrives at the following relation,

\begin{equation}
\frac{dV}{d\chi} =
\frac{3 \lambda}{2 f g^2} H^4 m_Q^3
\left[ 10 + 6 m_Q^2  
+ \frac{4 f^2 g^2}{m_Q^4 H^2 \lambda^2} + O(\varepsilon,\varepsilon^2, \dot{Q},\dot{Q}^2) \right]\; ,
\label{etaSBR}
\end{equation}
 which does not explicitly depend on backreaction terms. In deriving this equation, it is important to recall that all the relevant modes during the transition to the strong backreaction regime are super-horizon and behave as $\sqrt{2k}\, x\, t_L\sim c(k) $. Note that in Eq.~(\ref{etaSBR}) we keep terms up to order $1 / m_Q$, with $m_Q\gg1$.

Equipped with the above this relation, one can derive an expression for $\eta_\chi$ using the following definition:
\begin{equation}
    \eta_\chi \equiv \frac{V_{\chi\chi}}{3 H^2}  = \frac{1}{3 H^2} \dot{\chi}^{-1} \frac{dV_\chi}{dt}\; ,
\end{equation}
which gives:
\begin{align}
\eta_\chi & =
\frac{1}{4 f^2 g^2} 
\Bigg[ 12 f^2 g^2 \varepsilon  + \lambda^2 H^2 m_Q^4\varepsilon \left(10 -6m_Q^2 + (11 + 4 m_Q^2 )\frac{\dot{\varepsilon}}{H \varepsilon} \right) 
\nonumber \\ &\hspace{1cm} 
+
\left(4 f^2 g^2 
-30 \lambda^2 H^2 m_Q^4 (1 + m_Q^2 )  \right)\left(\frac{\dot{Q}}{H Q}\right)  
-2 \lambda^2 H^2 m_Q^4 (11 + 4 m_Q^2) \left( \frac{\dot{Q}}{H Q}\right)^2
\nonumber\\ &\hspace{10cm}
- 3 H^2 m_Q^4 \lambda^2 \left(\frac{\dot{Q}}{H Q}\right)^3  
 \Bigg].
\end{align}
Expanding only the highest power in $m_Q$ one finds the leading expression for $\eta_\chi$  at the end of inflation in the SB regime:
\begin{equation}
\eta_\chi  \approx
-\frac{1}{2} m_Q^4 \Lambda^2 
\left[ 3\varepsilon - 2\frac{\dot{\varepsilon}}{H }
+ 15  \left( \frac{\dot{Q}}{H Q}\right)
+4 \left( \frac{\dot{Q}}{H Q}\right)^2 \right].
\end{equation}

\newpage
\addcontentsline{toc}{section}{References}
\bibliographystyle{utphys}
\bibliography{paper2}

\providecommand{\href}[2]{#2}\begingroup\raggedright\begin{thebibliography}{100}

\bibitem{Brandenberger:2009jq}
R.~H. Brandenberger, ``{Alternatives to the inflationary paradigm of structure formation},'' \href{http://dx.doi.org/10.1142/S2010194511000109}{{\em Int. J. Mod. Phys. Conf. Ser.} {\bfseries 01} (2011) 67--79}, \href{http://arxiv.org/abs/0902.4731}{{\ttfamily arXiv:0902.4731 [hep-th]}}.

\bibitem{Lyth:1998xn}
D.~H. Lyth and A.~Riotto, ``{Particle physics models of inflation and the cosmological density perturbation},'' \href{http://dx.doi.org/10.1016/S0370-1573(98)00128-8}{{\em Phys. Rept.} {\bfseries 314} (1999) 1--146}, \href{http://arxiv.org/abs/hep-ph/9807278}{{\ttfamily arXiv:hep-ph/9807278}}.

\bibitem{Bird:2016dcv}
S.~Bird, I.~Cholis, J.~B. Muñoz, Y.~Ali-Haïmoud, M.~Kamionkowski, E.~D. Kovetz, A.~Raccanelli, and A.~G. Riess, ``{Did LIGO detect dark matter?},'' \href{http://dx.doi.org/10.1103/PhysRevLett.116.201301}{{\em Phys. Rev. Lett.} {\bfseries 116} no.~20, (2016) 201301},
\href{http://arxiv.org/abs/1603.00464}{{\ttfamily arXiv:1603.00464 [astro-ph.CO]}}.

\bibitem{Baumann:2014nda}
D.~Baumann and L.~McAllister, \href{http://dx.doi.org/10.1017/CBO9781316105733}{{\em {Inflation and String Theory}}}.
\newblock Cambridge Monographs on Mathematical Physics. Cambridge University Press, 5, 2015.
\newblock \href{http://arxiv.org/abs/1404.2601}{{\ttfamily arXiv:1404.2601 [hep-th]}}.

\bibitem{SimonsObservatory:2018koc}
{\bfseries Simons Observatory} Collaboration, P.~Ade {\em et~al.}, ``{The Simons Observatory: Science goals and forecasts},'' \href{http://dx.doi.org/10.1088/1475-7516/2019/02/056}{{\em JCAP} {\bfseries 02} (2019) 056}, \href{http://arxiv.org/abs/1808.07445}{{\ttfamily arXiv:1808.07445 [astro-ph.CO]}}.

\bibitem{CMB-S4:2016ple}
{\bfseries CMB-S4} Collaboration, K.~N. Abazajian {\em et~al.}, ``{CMB-S4 Science Book, First Edition},'' \href{http://arxiv.org/abs/1610.02743}{{\ttfamily arXiv:1610.02743 [astro-ph.CO]}}.

\bibitem{CMB-S4:2022ght}
{\bfseries CMB-S4} Collaboration, K.~Abazajian {\em et~al.}, ``{Snowmass 2021 CMB-S4 White Paper},'' \href{http://arxiv.org/abs/2203.08024}{{\ttfamily arXiv:2203.08024 [astro-ph.CO]}}.

\bibitem{Amendola:2016saw}
L.~Amendola {\em et~al.}, ``{Cosmology and fundamental physics with the Euclid satellite},'' \href{http://dx.doi.org/10.1007/s41114-017-0010-3}{{\em Living Rev. Rel.} {\bfseries 21} no.~1, (2018) 2}, \href{http://arxiv.org/abs/1606.00180}{{\ttfamily arXiv:1606.00180 [astro-ph.CO]}}.

\bibitem{LSSTScience:2009jmu}
{\bfseries LSST Science, LSST Project} Collaboration, P.~A. Abell {\em et~al.}, ``{LSST Science Book, Version 2.0},'' \href{http://arxiv.org/abs/0912.0201}{{\ttfamily arXiv:0912.0201 [astro-ph.IM]}}.

\bibitem{EPTA:2023xxk}
{\bfseries EPTA} Collaboration, J.~Antoniadis {\em et~al.}, ``{The second data release from the European Pulsar Timing Array: IV. Implications for massive black holes, dark matter and the early Universe},'' \href{http://arxiv.org/abs/2306.16227}{{\ttfamily arXiv:2306.16227 [astro-ph.CO]}}.

\bibitem{NANOGrav:2023gor}
{\bfseries NANOGrav} Collaboration, G.~Agazie {\em et~al.}, ``{The NANOGrav 15 yr Data Set: Evidence for a Gravitational-wave Background},'' \href{http://dx.doi.org/10.3847/2041-8213/acdac6}{{\em Astrophys. J. Lett.} {\bfseries 951} no.~1, (2023) L8}, \href{http://arxiv.org/abs/2306.16213}{{\ttfamily arXiv:2306.16213 [astro-ph.HE]}}.

\bibitem{SKA:2018ckk}
{\bfseries SKA} Collaboration, D.~J. Bacon {\em et~al.}, ``{Cosmology with Phase 1 of the Square Kilometre Array: Red Book 2018: Technical specifications and performance forecasts},'' \href{http://dx.doi.org/10.1017/pasa.2019.51}{{\em Publ. Astron. Soc. Austral.} {\bfseries 37} (2020) e007}, \href{http://arxiv.org/abs/1811.02743}{{\ttfamily arXiv:1811.02743 [astro-ph.CO]}}.

\bibitem{LIGOScientific:2016wof}
{\bfseries LIGO Scientific} Collaboration, B.~P. Abbott {\em et~al.}, ``{Exploring the Sensitivity of Next Generation Gravitational Wave Detectors},'' \href{http://dx.doi.org/10.1088/1361-6382/aa51f4}{{\em Class. Quant. Grav.} {\bfseries 34} no.~4, (2017) 044001}, \href{http://arxiv.org/abs/1607.08697}{{\ttfamily arXiv:1607.08697 [astro-ph.IM]}}.

\bibitem{LIGOScientific:2021aug}
{\bfseries LIGO Scientific, Virgo, KAGRA} Collaboration, R.~Abbott {\em et~al.}, ``{Constraints on the Cosmic Expansion History from GWTC\textendash{}3},'' \href{http://dx.doi.org/10.3847/1538-4357/ac74bb}{{\em Astrophys. J.} {\bfseries 949} no.~2, (2023) 76}, \href{http://arxiv.org/abs/2111.03604}{{\ttfamily arXiv:2111.03604 [astro-ph.CO]}}.

\bibitem{LISACosmologyWorkingGroup:2022jok}
{\bfseries LISA Cosmology Working Group} Collaboration, P.~Auclair {\em et~al.}, ``{Cosmology with the Laser Interferometer Space Antenna},'' \href{http://dx.doi.org/10.1007/s41114-023-00045-2}{{\em Living Rev. Rel.} {\bfseries 26} no.~1, (2023) 5}, \href{http://arxiv.org/abs/2204.05434}{{\ttfamily arXiv:2204.05434 [astro-ph.CO]}}.

\bibitem{Maggiore:2019uih}
M.~Maggiore {\em et~al.}, ``{Science Case for the Einstein Telescope},'' \href{http://dx.doi.org/10.1088/1475-7516/2020/03/050}{{\em JCAP} {\bfseries 03} (2020) 050}, \href{http://arxiv.org/abs/1912.02622}{{\ttfamily arXiv:1912.02622 [astro-ph.CO]}}.

\bibitem{Reitze:2019iox}
D.~Reitze {\em et~al.}, ``{Cosmic Explorer: The U.S. Contribution to Gravitational-Wave Astronomy beyond LIGO},'' {\em Bull. Am. Astron. Soc.} {\bfseries 51} no.~7, (2019) 035, \href{http://arxiv.org/abs/1907.04833}{{\ttfamily arXiv:1907.04833 [astro-ph.IM]}}.

\bibitem{Kawamura:2020pcg}
S.~Kawamura {\em et~al.}, ``{Current status of space gravitational wave antenna DECIGO and B-DECIGO},'' \href{http://dx.doi.org/10.1093/ptep/ptab019}{{\em PTEP} {\bfseries 2021} no.~5, (2021) 05A105}, \href{http://arxiv.org/abs/2006.13545}{{\ttfamily arXiv:2006.13545 [gr-qc]}}.

\bibitem{Munoz:2015eqa}
J.~B. Mu\~noz, Y.~Ali-Ha\"\i{}moud, and M.~Kamionkowski, ``{Primordial non-gaussianity from the bispectrum of 21-cm fluctuations in the dark ages},'' \href{http://dx.doi.org/10.1103/PhysRevD.92.083508}{{\em Phys. Rev. D} {\bfseries 92} no.~8, (2015) 083508}, \href{http://arxiv.org/abs/1506.04152}{{\ttfamily arXiv:1506.04152 [astro-ph.CO]}}.

\bibitem{Freese:1990rb}
K.~Freese, J.~A. Frieman, and A.~V. Olinto, ``{Natural inflation with pseudo - Nambu-Goldstone bosons},''
\href{http://dx.doi.org/10.1103/PhysRevLett.65.3233}{{\em Phys.Rev.Lett.} {\bfseries 65} (1990) 3233--3236}.

\bibitem{Adams:1992bn}
F.~C. Adams, J.~R. Bond, K.~Freese, J.~A. Frieman, and A.~V. Olinto, ``{Natural inflation: Particle physics models, power law spectra for large scale structure, and constraints from COBE},'' \href{http://dx.doi.org/10.1103/PhysRevD.47.426}{{\em Phys.Rev.} {\bfseries D47} (1993) 426--455},
\href{http://arxiv.org/abs/hep-ph/9207245}{{\ttfamily arXiv:hep-ph/9207245 [hep-ph]}}.

\bibitem{Freese:2014nla}
K.~Freese and W.~H. Kinney, ``{Natural Inflation: Consistency with Cosmic Microwave Background Observations of Planck and BICEP2},'' \href{http://dx.doi.org/10.1088/1475-7516/2015/03/044}{{\em JCAP} {\bfseries 03} (2015) 044}, \href{http://arxiv.org/abs/1403.5277}{{\ttfamily arXiv:1403.5277 [astro-ph.CO]}}.

\bibitem{DallAgata:2018ybl}
G.~Dall'Agata, ``{Chromo-Natural inflation in Supergravity},'' \href{http://dx.doi.org/10.1016/j.physletb.2018.05.020}{{\em Phys. Lett. B} {\bfseries 782} (2018) 139--142}, \href{http://arxiv.org/abs/1804.03104}{{\ttfamily arXiv:1804.03104 [hep-th]}}.

\bibitem{Holland:2020jdh}
J.~Holland, I.~Zavala, and G.~Tasinato, ``{On chromonatural inflation in string theory},'' \href{http://dx.doi.org/10.1088/1475-7516/2020/12/026}{{\em JCAP} {\bfseries 12} (2020) 026}, \href{http://arxiv.org/abs/2009.00653}{{\ttfamily arXiv:2009.00653 [hep-th]}}.

\bibitem{Dimastrogiovanni:2023juq}
E.~Dimastrogiovanni, M.~Fasiello, J.~M. Leedom, M.~Putti, and A.~Westphal, ``{Gravitational Axiverse Spectroscopy: Seeing the Forest for the Axions},'' \href{http://arxiv.org/abs/2312.13431}{{\ttfamily arXiv:2312.13431 [hep-th]}}.

\bibitem{Nomura:2017ehb}
Y.~Nomura, T.~Watari, and M.~Yamazaki, ``{Pure Natural Inflation},'' \href{http://dx.doi.org/10.1016/j.physletb.2017.11.052}{{\em Phys. Lett. B} {\bfseries 776} (2018) 227--230}, \href{http://arxiv.org/abs/1706.08522}{{\ttfamily arXiv:1706.08522 [hep-ph]}}.

\bibitem{Nomura:2017zqj}
Y.~Nomura and M.~Yamazaki, ``{Tensor Modes in Pure Natural Inflation},'' \href{http://dx.doi.org/10.1016/j.physletb.2018.02.071}{{\em Phys. Lett. B} {\bfseries 780} (2018) 106--110}, \href{http://arxiv.org/abs/1711.10490}{{\ttfamily arXiv:1711.10490 [hep-ph]}}.

\bibitem{Adshead:2012kp}
P.~Adshead and M.~Wyman, ``{Chromo-Natural Inflation: Natural inflation on a steep potential with classical non-Abelian gauge fields},'' \href{http://dx.doi.org/10.1103/PhysRevLett.108.261302}{{\em Phys. Rev. Lett.} {\bfseries 108} (2012) 261302}, \href{http://arxiv.org/abs/1202.2366}{{\ttfamily arXiv:1202.2366 [hep-th]}}.

\bibitem{Dimastrogiovanni:2012ew}
E.~Dimastrogiovanni and M.~Peloso, ``{Stability analysis of chromo-natural inflation and possible evasion of Lyth's bound},'' \href{http://dx.doi.org/10.1103/PhysRevD.87.103501}{{\em Phys. Rev.} {\bfseries D87} no.~10, (2013) 103501},
\href{http://arxiv.org/abs/1212.5184}{{\ttfamily arXiv:1212.5184 [astro-ph.CO]}}.

\bibitem{Dimastrogiovanni:2012st}
E.~Dimastrogiovanni, M.~Fasiello, and A.~J. Tolley, ``{Low-Energy Effective Field Theory for Chromo-Natural Inflation},'' \href{http://dx.doi.org/10.1088/1475-7516/2013/02/046}{{\em JCAP} {\bfseries 02} (2013) 046}, \href{http://arxiv.org/abs/1211.1396}{{\ttfamily arXiv:1211.1396 [hep-th]}}.

\bibitem{Maleknejad:2012fw}
A.~Maleknejad, M.~M. Sheikh-Jabbari, and J.~Soda, ``{Gauge Fields and Inflation},'' \href{http://dx.doi.org/10.1016/j.physrep.2013.03.003}{{\em Phys. Rept.} {\bfseries 528} (2013) 161--261}, \href{http://arxiv.org/abs/1212.2921}{{\ttfamily arXiv:1212.2921 [hep-th]}}.

\bibitem{Adshead:2013nka}
P.~Adshead, E.~Martinec, and M.~Wyman, ``{Perturbations in Chromo-Natural Inflation},'' \href{http://dx.doi.org/10.1007/JHEP09(2013)087}{{\em JHEP} {\bfseries 09} (2013) 087}, \href{http://arxiv.org/abs/1305.2930}{{\ttfamily arXiv:1305.2930 [hep-th]}}.

\bibitem{Maleknejad:2013npa}
A.~Maleknejad and E.~Erfani, ``{Chromo-Natural Model in Anisotropic Background},'' \href{http://dx.doi.org/10.1088/1475-7516/2014/03/016}{{\em JCAP} {\bfseries 03} (2014) 016}, \href{http://arxiv.org/abs/1311.3361}{{\ttfamily arXiv:1311.3361 [hep-th]}}.

\bibitem{Maleknejad:2014wsa}
A.~Maleknejad, ``{Chiral Gravity Waves and Leptogenesis in Inflationary Models with non-Abelian Gauge Fields},'' \href{http://dx.doi.org/10.1103/PhysRevD.90.023542}{{\em Phys. Rev. D} {\bfseries 90} no.~2, (2014) 023542}, \href{http://arxiv.org/abs/1401.7628}{{\ttfamily arXiv:1401.7628 [hep-th]}}.

\bibitem{Adshead:2016omu}
P.~Adshead, E.~Martinec, E.~I. Sfakianakis, and M.~Wyman, ``{Higgsed Chromo-Natural Inflation},'' \href{http://dx.doi.org/10.1007/JHEP12(2016)137}{{\em JHEP} {\bfseries 12} (2016) 137}, \href{http://arxiv.org/abs/1609.04025}{{\ttfamily arXiv:1609.04025 [hep-th]}}.

\bibitem{Maleknejad:2016dci}
A.~Maleknejad, ``{Gravitational leptogenesis in axion inflation with SU(2) gauge field},'' \href{http://dx.doi.org/10.1088/1475-7516/2016/12/027}{{\em JCAP} {\bfseries 12} (2016) 027}, \href{http://arxiv.org/abs/1604.06520}{{\ttfamily arXiv:1604.06520 [hep-ph]}}.

\bibitem{Maleknejad:2018nxz}
A.~Maleknejad and E.~Komatsu, ``{Production and Backreaction of Spin-2 Particles of $SU(2)$ Gauge Field during Inflation},'' \href{http://dx.doi.org/10.1007/JHEP05(2019)174}{{\em JHEP} {\bfseries 05} (2019) 174}, \href{http://arxiv.org/abs/1808.09076}{{\ttfamily arXiv:1808.09076 [hep-ph]}}.

\bibitem{Papageorgiou:2018rfx}
A.~Papageorgiou, M.~Peloso, and C.~Unal, ``{Nonlinear perturbations from the coupling of the inflaton to a non-Abelian gauge field, with a focus on Chromo-Natural Inflation},'' \href{http://dx.doi.org/10.1088/1475-7516/2018/09/030}{{\em JCAP} {\bfseries 09} (2018) 030}, \href{http://arxiv.org/abs/1806.08313}{{\ttfamily arXiv:1806.08313 [astro-ph.CO]}}.

\bibitem{Lozanov:2018kpk}
K.~D. Lozanov, A.~Maleknejad, and E.~Komatsu, ``{Schwinger Effect by an $SU(2)$ Gauge Field during Inflation},'' \href{http://dx.doi.org/10.1007/JHEP02(2019)041}{{\em JHEP} {\bfseries 02} (2019) 041}, \href{http://arxiv.org/abs/1805.09318}{{\ttfamily arXiv:1805.09318 [hep-th]}}.

\bibitem{Agrawal:2018mkd}
P.~Agrawal, J.~Fan, and M.~Reece, ``{Clockwork Axions in Cosmology: Is Chromonatural Inflation Chrononatural?},'' \href{http://dx.doi.org/10.1007/JHEP10(2018)193}{{\em JHEP} {\bfseries 10} (2018) 193}, \href{http://arxiv.org/abs/1806.09621}{{\ttfamily arXiv:1806.09621 [hep-th]}}.

\bibitem{Mirzagholi:2019jeb}
L.~Mirzagholi, A.~Maleknejad, and K.~D. Lozanov, ``{Production and backreaction of fermions from axion-$SU(2)$ gauge fields during inflation},'' \href{http://dx.doi.org/10.1103/PhysRevD.101.083528}{{\em Phys. Rev. D} {\bfseries 101} no.~8, (2020) 083528}, \href{http://arxiv.org/abs/1905.09258}{{\ttfamily arXiv:1905.09258 [hep-th]}}.

\bibitem{Fujita:2022fff}
T.~Fujita, K.~Murai, and R.~Namba, ``{Universality of linear perturbations in SU(N) natural inflation},'' \href{http://dx.doi.org/10.1103/PhysRevD.105.103518}{{\em Phys. Rev. D} {\bfseries 105} no.~10, (2022) 103518}, \href{http://arxiv.org/abs/2203.03977}{{\ttfamily arXiv:2203.03977 [hep-ph]}}.

\bibitem{Dimastrogiovanni:2023oid}
E.~Dimastrogiovanni, M.~Fasiello, M.~Michelotti, and L.~Pinol, ``{Primordial gravitational waves in non-minimally coupled chromo-natural inflation},'' \href{http://dx.doi.org/10.1088/1475-7516/2024/02/039}{{\em JCAP} {\bfseries 02} (2024) 039}, \href{http://arxiv.org/abs/2303.10718}{{\ttfamily arXiv:2303.10718 [astro-ph.CO]}}.

\bibitem{Murata:2024urv}
T.~Murata and T.~Kobayashi, ``{Chromo-natural inflation supported by enhanced friction from Horndeski gravity},'' \href{http://dx.doi.org/10.1088/1475-7516/2024/10/044}{{\em JCAP} {\bfseries 10} (2024) 044}, \href{http://arxiv.org/abs/2408.01773}{{\ttfamily arXiv:2408.01773 [astro-ph.CO]}}.

\bibitem{Michelotti:2024bbc}
M.~Michelotti, R.~Gonzalez~Quaglia, E.~Dimastrogiovanni, M.~Fasiello, and D.~Roest, ``{Kinetic Gauge Friction in Natural Inflation},'' \href{http://arxiv.org/abs/2411.19892}{{\ttfamily arXiv:2411.19892 [astro-ph.CO]}}.

\bibitem{Galtsov:1991un}
D.~V. Galtsov and M.~S. Volkov, ``{Yang-Mills cosmology: Cold matter for a hot universe},'' \href{http://dx.doi.org/10.1016/0370-2693(91)90211-8}{{\em Phys. Lett. B} {\bfseries 256} (1991) 17--21}.

\bibitem{Maleknejad:2011jw}
A.~Maleknejad and M.~M. Sheikh-Jabbari, ``{Gauge-flation: Inflation From Non-Abelian Gauge Fields},'' \href{http://dx.doi.org/10.1016/j.physletb.2013.05.001}{{\em Phys. Lett. B} {\bfseries 723} (2013) 224--228}, \href{http://arxiv.org/abs/1102.1513}{{\ttfamily arXiv:1102.1513 [hep-ph]}}.

\bibitem{Maleknejad:2011jr}
A.~Maleknejad, M.~M. Sheikh-Jabbari, and J.~Soda, ``{Gauge-flation and Cosmic No-Hair Conjecture},'' \href{http://dx.doi.org/10.1088/1475-7516/2012/01/016}{{\em JCAP} {\bfseries 01} (2012) 016}, \href{http://arxiv.org/abs/1109.5573}{{\ttfamily arXiv:1109.5573 [hep-th]}}.

\bibitem{Wolfson:2020fqz}
I.~Wolfson, A.~Maleknejad, and E.~Komatsu, ``{How attractive is the isotropic attractor solution of axion-SU(2) inflation?},'' \href{http://dx.doi.org/10.1088/1475-7516/2020/09/047}{{\em JCAP} {\bfseries 09} (2020) 047}, \href{http://arxiv.org/abs/2003.01617}{{\ttfamily arXiv:2003.01617 [gr-qc]}}.

\bibitem{Domcke:2018rvv}
V.~Domcke, B.~Mares, F.~Muia, and M.~Pieroni, ``{Emerging chromo-natural inflation},'' \href{http://dx.doi.org/10.1088/1475-7516/2019/04/034}{{\em JCAP} {\bfseries 04} (2019) 034}, \href{http://arxiv.org/abs/1807.03358}{{\ttfamily arXiv:1807.03358 [hep-ph]}}.

\bibitem{Domcke:2019lxq}
V.~Domcke and S.~Sandner, ``{The Different Regimes of Axion Gauge Field Inflation},'' \href{http://dx.doi.org/10.1088/1475-7516/2019/09/038}{{\em JCAP} {\bfseries 09} (2019) 038}, \href{http://arxiv.org/abs/1905.11372}{{\ttfamily arXiv:1905.11372 [astro-ph.CO]}}.

\bibitem{Papageorgiou:2019ecb}
A.~Papageorgiou, M.~Peloso, and C.~Unal, ``{Nonlinear perturbations from axion-gauge fields dynamics during inflation},'' \href{http://dx.doi.org/10.1088/1475-7516/2019/07/004}{{\em JCAP} {\bfseries 07} (2019) 004}, \href{http://arxiv.org/abs/1904.01488}{{\ttfamily arXiv:1904.01488 [astro-ph.CO]}}.

\bibitem{Badger:2024ekb}
C.~Badger, H.~Duval, T.~Fujita, S.~Kuroyanagi, A.~Romero-Rodr\'\i{}guez, and M.~Sakellariadou, ``{Detection prospects of gravitational waves from SU(2) axion inflation},'' \href{http://dx.doi.org/10.1103/PhysRevD.110.084063}{{\em Phys. Rev. D} {\bfseries 110} no.~8, (2024) 084063}, \href{http://arxiv.org/abs/2406.11742}{{\ttfamily arXiv:2406.11742 [astro-ph.CO]}}.

\bibitem{Seto:2007tn}
N.~Seto and A.~Taruya, ``{Measuring a Parity Violation Signature in the Early Universe via Ground-based Laser Interferometers},'' \href{http://dx.doi.org/10.1103/PhysRevLett.99.121101}{{\em Phys. Rev. Lett.} {\bfseries 99} (2007) 121101}, \href{http://arxiv.org/abs/0707.0535}{{\ttfamily arXiv:0707.0535 [astro-ph]}}.

\bibitem{Smith:2016jqs}
T.~L. Smith and R.~Caldwell, ``{Sensitivity to a Frequency-Dependent Circular Polarization in an Isotropic Stochastic Gravitational Wave Background},'' \href{http://dx.doi.org/10.1103/PhysRevD.95.044036}{{\em Phys. Rev. D} {\bfseries 95} no.~4, (2017) 044036}, \href{http://arxiv.org/abs/1609.05901}{{\ttfamily arXiv:1609.05901 [gr-qc]}}.

\bibitem{Domcke:2019zls}
V.~Domcke, J.~Garcia-Bellido, M.~Peloso, M.~Pieroni, A.~Ricciardone, L.~Sorbo, and G.~Tasinato, ``{Measuring the net circular polarization of the stochastic gravitational wave background with interferometers},'' \href{http://dx.doi.org/10.1088/1475-7516/2020/05/028}{{\em JCAP} {\bfseries 05} (2020) 028}, \href{http://arxiv.org/abs/1910.08052}{{\ttfamily arXiv:1910.08052 [astro-ph.CO]}}.

\bibitem{Unal:2023srk}
C.~Unal, A.~Papageorgiou, and I.~Obata, ``{Axion-gauge dynamics during inflation as the origin of pulsar timing array signals and primordial black holes},'' \href{http://dx.doi.org/10.1016/j.physletb.2024.138873}{{\em Phys. Lett. B} {\bfseries 856} (2024) 138873}, \href{http://arxiv.org/abs/2307.02322}{{\ttfamily arXiv:2307.02322 [astro-ph.CO]}}.

\bibitem{Anber:2009ua}
M.~M. Anber and L.~Sorbo, ``{Naturally inflating on steep potentials through electromagnetic dissipation},'' \href{http://dx.doi.org/10.1103/PhysRevD.81.043534}{{\em Phys. Rev.} {\bfseries D81} (2010) 043534},
\href{http://arxiv.org/abs/0908.4089}{{\ttfamily arXiv:0908.4089 [hep-th]}}.

\bibitem{Peloso:2022ovc}
M.~Peloso and L.~Sorbo, ``{Instability in axion inflation with strong backreaction from gauge modes},'' \href{http://arxiv.org/abs/2209.08131}{{\ttfamily arXiv:2209.08131 [astro-ph.CO]}}.

\bibitem{vonEckardstein:2023gwk}
R.~von Eckardstein, M.~Peloso, K.~Schmitz, O.~Sobol, and L.~Sorbo, ``{Axion inflation in the strong-backreaction regime: decay of the Anber-Sorbo solution},'' \href{http://dx.doi.org/10.1007/JHEP11(2023)183}{{\em JHEP} {\bfseries 11} (2023) 183}, \href{http://arxiv.org/abs/2309.04254}{{\ttfamily arXiv:2309.04254 [hep-ph]}}.

\bibitem{Caravano:2022epk}
A.~Caravano, E.~Komatsu, K.~D. Lozanov, and J.~Weller, ``{Lattice simulations of axion-U(1) inflation},'' \href{http://dx.doi.org/10.1103/PhysRevD.108.043504}{{\em Phys. Rev. D} {\bfseries 108} no.~4, (2023) 043504}, \href{http://arxiv.org/abs/2204.12874}{{\ttfamily arXiv:2204.12874 [astro-ph.CO]}}.

\bibitem{Figueroa:2023oxc}
D.~G. Figueroa, J.~Lizarraga, A.~Urio, and J.~Urrestilla, ``{Strong Backreaction Regime in Axion Inflation},'' \href{http://dx.doi.org/10.1103/PhysRevLett.131.151003}{{\em Phys. Rev. Lett.} {\bfseries 131} no.~15, (2023) 151003}, \href{http://arxiv.org/abs/2303.17436}{{\ttfamily arXiv:2303.17436 [astro-ph.CO]}}.

\bibitem{Caravano:2024xsb}
A.~Caravano and M.~Peloso, ``{Unveiling the nonlinear dynamics of a rolling axion during inflation},'' \href{http://arxiv.org/abs/2407.13405}{{\ttfamily arXiv:2407.13405 [astro-ph.CO]}}.

\bibitem{Figueroa:2024rkr}
D.~G. Figueroa, J.~Lizarraga, N.~Loayza, A.~Urio, and J.~Urrestilla, ``{Nonlinear dynamics of axion inflation: A detailed lattice study},'' \href{http://dx.doi.org/10.1103/PhysRevD.111.063545}{{\em Phys. Rev. D} {\bfseries 111} no.~6, (2025) 063545}, \href{http://arxiv.org/abs/2411.16368}{{\ttfamily arXiv:2411.16368 [astro-ph.CO]}}.

\bibitem{Sharma:2024nfu}
R.~Sharma, A.~Brandenburg, K.~Subramanian, and A.~Vikman, ``{Lattice simulations of axion-U(1) inflation: gravitational waves, magnetic fields, and scalar statistics},'' \href{http://arxiv.org/abs/2411.04854}{{\ttfamily arXiv:2411.04854 [astro-ph.CO]}}.

\bibitem{Iarygina:2023mtj}
O.~Iarygina, E.~I. Sfakianakis, R.~Sharma, and A.~Brandenburg, ``{Backreaction of axion-SU(2) dynamics during inflation},'' \href{http://dx.doi.org/10.1088/1475-7516/2024/04/018}{{\em JCAP} {\bfseries 04} (2024) 018}, \href{http://arxiv.org/abs/2311.07557}{{\ttfamily arXiv:2311.07557 [astro-ph.CO]}}.

\bibitem{Dimastrogiovanni:2024xvc}
E.~Dimastrogiovanni, M.~Fasiello, and A.~Papageorgiou, ``{A novel PBH production mechanism from non-Abelian gauge fields during inflation},'' \href{http://arxiv.org/abs/2403.13581}{{\ttfamily arXiv:2403.13581 [astro-ph.CO]}}.

\bibitem{Barnaby:2012xt}
N.~Barnaby, J.~Moxon, R.~Namba, M.~Peloso, G.~Shiu, {\em et~al.}, ``{Gravity waves and non-Gaussian features from particle production in a sector gravitationally coupled to the inflaton},'' \href{http://dx.doi.org/10.1103/PhysRevD.86.103508}{{\em Phys.Rev.} {\bfseries D86} (2012) 103508},
\href{http://arxiv.org/abs/1206.6117}{{\ttfamily arXiv:1206.6117 [astro-ph.CO]}}.

\bibitem{Mukohyama:2014gba}
S.~Mukohyama, R.~Namba, M.~Peloso, and G.~Shiu, ``{Blue Tensor Spectrum from Particle Production during Inflation},'' \href{http://dx.doi.org/10.1088/1475-7516/2014/08/036}{{\em JCAP} {\bfseries 08} (2014) 036}, \href{http://arxiv.org/abs/1405.0346}{{\ttfamily arXiv:1405.0346 [astro-ph.CO]}}.

\bibitem{Namba:2015gja}
R.~Namba, M.~Peloso, M.~Shiraishi, L.~Sorbo, and C.~Unal, ``{Scale-dependent gravitational waves from a rolling axion},'' \href{http://dx.doi.org/10.1088/1475-7516/2016/01/041}{{\em JCAP} {\bfseries 1601} no.~01, (2016) 041},
\href{http://arxiv.org/abs/1509.07521}{{\ttfamily arXiv:1509.07521 [astro-ph.CO]}}.

\bibitem{Garcia-Bellido:2016dkw}
J.~Garcia-Bellido, M.~Peloso, and C.~Unal, ``{Gravitational waves at interferometer scales and primordial black holes in axion inflation},'' \href{http://dx.doi.org/10.1088/1475-7516/2016/12/031}{{\em JCAP} {\bfseries 1612} no.~12, (2016) 031},
\href{http://arxiv.org/abs/1610.03763}{{\ttfamily arXiv:1610.03763 [astro-ph.CO]}}.

\bibitem{Obata:2016tmo}
{\bfseries CLEO} Collaboration, I.~Obata and J.~Soda, ``{Chiral primordial Chiral primordial gravitational waves from dilaton induced delayed chromonatural inflation},'' \href{http://dx.doi.org/10.1103/PhysRevD.95.109903, 10.1103/PhysRevD.93.123502}{{\em Phys. Rev.} {\bfseries D93} no.~12, (2016) 123502}, \href{http://arxiv.org/abs/1602.06024}{{\ttfamily arXiv:1602.06024 [hep-th]}}.
[Addendum: Phys. Rev.D95,no.10,109903(2017)].

\bibitem{Dimastrogiovanni:2016fuu}
E.~Dimastrogiovanni, M.~Fasiello, and T.~Fujita, ``{Primordial Gravitational Waves from Axion-Gauge Fields Dynamics},'' \href{http://dx.doi.org/10.1088/1475-7516/2017/01/019}{{\em JCAP} {\bfseries 01} (2017) 019}, \href{http://arxiv.org/abs/1608.04216}{{\ttfamily arXiv:1608.04216 [astro-ph.CO]}}.

\bibitem{Agrawal:2017awz}
A.~Agrawal, T.~Fujita, and E.~Komatsu, ``{Large tensor non-Gaussianity from axion-gauge field dynamics},'' \href{http://dx.doi.org/10.1103/PhysRevD.97.103526}{{\em Phys. Rev. D} {\bfseries 97} no.~10, (2018) 103526}, \href{http://arxiv.org/abs/1707.03023}{{\ttfamily arXiv:1707.03023 [astro-ph.CO]}}.

\bibitem{Thorne:2017jft}
B.~Thorne, T.~Fujita, M.~Hazumi, N.~Katayama, E.~Komatsu, and M.~Shiraishi, ``{Finding the chiral gravitational wave background of an axion-SU(2) inflationary model using CMB observations and laser interferometers},'' \href{http://dx.doi.org/10.1103/PhysRevD.97.043506}{{\em Phys. Rev. D} {\bfseries 97} no.~4, (2018) 043506}, \href{http://arxiv.org/abs/1707.03240}{{\ttfamily arXiv:1707.03240 [astro-ph.CO]}}.

\bibitem{Fujita:2017jwq}
T.~Fujita, R.~Namba, and Y.~Tada, ``{Does the detection of primordial gravitational waves exclude low energy inflation?},'' \href{http://dx.doi.org/10.1016/j.physletb.2017.12.014}{{\em Phys. Lett. B} {\bfseries 778} (2018) 17--21}, \href{http://arxiv.org/abs/1705.01533}{{\ttfamily arXiv:1705.01533 [astro-ph.CO]}}.

\bibitem{Agrawal:2018gzp}
A.~Agrawal, ``{Non-Gaussianity of Inflationary Gravitational Waves from the Field Equation},'' \href{http://dx.doi.org/10.1142/S0218271819500366}{{\em Int. J. Mod. Phys. D} {\bfseries 28} no.~02, (2018) 1950036}, \href{http://arxiv.org/abs/1804.01481}{{\ttfamily arXiv:1804.01481 [astro-ph.CO]}}.

\bibitem{Dimastrogiovanni:2018xnn}
E.~Dimastrogiovanni, M.~Fasiello, R.~J. Hardwick, H.~Assadullahi, K.~Koyama, and D.~Wands, ``{Non-Gaussianity from Axion-Gauge Fields Interactions during Inflation},'' \href{http://dx.doi.org/10.1088/1475-7516/2018/11/029}{{\em JCAP} {\bfseries 11} (2018) 029}, \href{http://arxiv.org/abs/1806.05474}{{\ttfamily arXiv:1806.05474 [astro-ph.CO]}}.

\bibitem{Fujita:2018vmv}
T.~Fujita, R.~Namba, and I.~Obata, ``{Mixed Non-Gaussianity from Axion-Gauge Field Dynamics},'' \href{http://dx.doi.org/10.1088/1475-7516/2019/04/044}{{\em JCAP} {\bfseries 04} (2019) 044}, \href{http://arxiv.org/abs/1811.12371}{{\ttfamily arXiv:1811.12371 [astro-ph.CO]}}.

\bibitem{Mirzagholi:2020irt}
L.~Mirzagholi, E.~Komatsu, K.~D. Lozanov, and Y.~Watanabe, ``{Effects of Gravitational Chern-Simons during Axion-SU(2) Inflation},'' \href{http://dx.doi.org/10.1088/1475-7516/2020/06/024}{{\em JCAP} {\bfseries 06} (2020) 024}, \href{http://arxiv.org/abs/2003.05931}{{\ttfamily arXiv:2003.05931 [gr-qc]}}.

\bibitem{Iarygina:2021bxq}
O.~Iarygina and E.~I. Sfakianakis, ``{Gravitational waves from spectator Gauge-flation},'' \href{http://dx.doi.org/10.1088/1475-7516/2021/11/023}{{\em JCAP} {\bfseries 11} no.~11, (2021) 023}, \href{http://arxiv.org/abs/2105.06972}{{\ttfamily arXiv:2105.06972 [hep-th]}}.

\bibitem{Fujita:2021flu}
T.~Fujita, K.~Murai, I.~Obata, and M.~Shiraishi, ``{Gravitational wave trispectrum in the axion-SU(2) model},'' \href{http://dx.doi.org/10.1088/1475-7516/2022/01/007}{{\em JCAP} {\bfseries 01} no.~01, (2022) 007}, \href{http://arxiv.org/abs/2109.06457}{{\ttfamily arXiv:2109.06457 [astro-ph.CO]}}.

\bibitem{Kakizaki:2021mgj}
M.~Kakizaki, M.~Ogata, and O.~Seto, ``{Dark radiation in spectator axion\textendash{}gauge models},'' \href{http://dx.doi.org/10.1093/ptep/ptac029}{{\em PTEP} {\bfseries 2022} no.~3, (2022) 033E02}, \href{http://arxiv.org/abs/2110.12936}{{\ttfamily arXiv:2110.12936 [hep-ph]}}.

\bibitem{Ishiwata:2021yne}
K.~Ishiwata, E.~Komatsu, and I.~Obata, ``{Axion-gauge field dynamics with backreaction},'' \href{http://dx.doi.org/10.1088/1475-7516/2022/03/010}{{\em JCAP} {\bfseries 03} no.~03, (2022) 010}, \href{http://arxiv.org/abs/2111.14429}{{\ttfamily arXiv:2111.14429 [hep-ph]}}.

\bibitem{Bagherian:2022mau}
H.~Bagherian, M.~Reece, and W.~L. Xu, ``{The inflated Chern-Simons number in spectator chromo-natural inflation},'' \href{http://dx.doi.org/10.1007/JHEP01(2023)099}{{\em JHEP} {\bfseries 01} (2023) 099}, \href{http://arxiv.org/abs/2207.11262}{{\ttfamily arXiv:2207.11262 [hep-th]}}.

\bibitem{Putti:2024uyr}
M.~Putti, N.~Bartolo, S.~Bhattacharya, and M.~Peloso, ``{CMB spectral distortions from enhanced primordial perturbations: the role of spectator axions},'' \href{http://arxiv.org/abs/2403.08594}{{\ttfamily arXiv:2403.08594 [astro-ph.CO]}}.

\bibitem{Brandenburg:2024awd}
A.~Brandenburg, O.~Iarygina, E.~I. Sfakianakis, and R.~Sharma, ``{Magnetogenesis from axion-SU(2) inflation},'' \href{http://dx.doi.org/10.1088/1475-7516/2024/12/057}{{\em JCAP} {\bfseries 12} (2024) 057}, \href{http://arxiv.org/abs/2408.17413}{{\ttfamily arXiv:2408.17413 [astro-ph.CO]}}.

\bibitem{Arvanitaki:2009fg}
A.~Arvanitaki, S.~Dimopoulos, S.~Dubovsky, N.~Kaloper, and J.~March-Russell, ``{String Axiverse},'' \href{http://dx.doi.org/10.1103/PhysRevD.81.123530}{{\em Phys. Rev. D} {\bfseries 81} (2010) 123530}, \href{http://arxiv.org/abs/0905.4720}{{\ttfamily arXiv:0905.4720 [hep-th]}}.

\bibitem{Acharya:2010zx}
B.~S. Acharya, K.~Bobkov, and P.~Kumar, ``{An M Theory Solution to the Strong CP Problem and Constraints on the Axiverse},'' \href{http://dx.doi.org/10.1007/JHEP11(2010)105}{{\em JHEP} {\bfseries 11} (2010) 105}, \href{http://arxiv.org/abs/1004.5138}{{\ttfamily arXiv:1004.5138 [hep-th]}}.

\bibitem{Cicoli:2012sz}
M.~Cicoli, M.~Goodsell, and A.~Ringwald, ``{The type IIB string axiverse and its low-energy phenomenology},'' \href{http://dx.doi.org/10.1007/JHEP10(2012)146}{{\em JHEP} {\bfseries 1210} (2012) 146},
\href{http://arxiv.org/abs/1206.0819}{{\ttfamily arXiv:1206.0819 [hep-th]}}.

\bibitem{Demirtas:2018akl}
M.~Demirtas, C.~Long, L.~McAllister, and M.~Stillman, ``{The Kreuzer-Skarke Axiverse},'' \href{http://dx.doi.org/10.1007/JHEP04(2020)138}{{\em JHEP} {\bfseries 04} (2020) 138}, \href{http://arxiv.org/abs/1808.01282}{{\ttfamily arXiv:1808.01282 [hep-th]}}.

\bibitem{Demirtas:2021gsq}
M.~Demirtas, N.~Gendler, C.~Long, L.~McAllister, and J.~Moritz, ``{PQ axiverse},'' \href{http://dx.doi.org/10.1007/JHEP06(2023)092}{{\em JHEP} {\bfseries 06} (2023) 092}, \href{http://arxiv.org/abs/2112.04503}{{\ttfamily arXiv:2112.04503 [hep-th]}}.

\bibitem{DAmico:2021vka}
G.~D'Amico, N.~Kaloper, and A.~Westphal, ``{Double Monodromy Inflation: A Gravity Waves Factory for CMB-S4, LiteBIRD and LISA},'' \href{http://dx.doi.org/10.1103/PhysRevD.104.L081302}{{\em Phys. Rev. D} {\bfseries 104} no.~8, (2021) L081302}, \href{http://arxiv.org/abs/2101.05861}{{\ttfamily arXiv:2101.05861 [hep-th]}}.

\bibitem{DAmico:2021fhz}
G.~D'Amico, N.~Kaloper, and A.~Westphal, ``{General double monodromy inflation},'' \href{http://dx.doi.org/10.1103/PhysRevD.105.103527}{{\em Phys. Rev. D} {\bfseries 105} no.~10, (2022) 103527}, \href{http://arxiv.org/abs/2112.13861}{{\ttfamily arXiv:2112.13861 [hep-th]}}.

\bibitem{Hashiba:2021gmn}
S.~Hashiba, K.~Kamada, and H.~Nakatsuka, ``{Gauge field production and Schwinger reheating in runaway axion inflation},'' \href{http://dx.doi.org/10.1088/1475-7516/2022/04/058}{{\em JCAP} {\bfseries 04} no.~04, (2022) 058}, \href{http://arxiv.org/abs/2110.10822}{{\ttfamily arXiv:2110.10822 [hep-ph]}}.

\bibitem{Spokoiny:1993kt}
B.~Spokoiny, ``{Deflationary universe scenario},'' \href{http://dx.doi.org/10.1016/0370-2693(93)90155-B}{{\em Phys. Lett. B} {\bfseries 315} (1993) 40--45}, \href{http://arxiv.org/abs/gr-qc/9306008}{{\ttfamily arXiv:gr-qc/9306008}}.

\bibitem{Peebles:1998qn}
P.~J.~E. Peebles and A.~Vilenkin, ``{Quintessential inflation},'' \href{http://dx.doi.org/10.1103/PhysRevD.59.063505}{{\em Phys. Rev. D} {\bfseries 59} (1999) 063505}, \href{http://arxiv.org/abs/astro-ph/9810509}{{\ttfamily arXiv:astro-ph/9810509}}.

\bibitem{tHooft:1973alw}
G.~'t~Hooft, ``{A Planar Diagram Theory for Strong Interactions},'' \href{http://dx.doi.org/10.1016/0550-3213(74)90154-0}{{\em Nucl. Phys. B} {\bfseries 72} (1974) 461}.

\bibitem{Witten:1980sp}
E.~Witten, ``{Large N Chiral Dynamics},'' \href{http://dx.doi.org/10.1016/0003-4916(80)90325-5}{{\em Annals Phys.} {\bfseries 128} (1980) 363}.

\bibitem{Witten:1998uka}
E.~Witten, ``{Theta dependence in the large N limit of four-dimensional gauge theories},'' \href{http://dx.doi.org/10.1103/PhysRevLett.81.2862}{{\em Phys. Rev. Lett.} {\bfseries 81} (1998) 2862--2865}, \href{http://arxiv.org/abs/hep-th/9807109}{{\ttfamily arXiv:hep-th/9807109}}.

\bibitem{Chatrchyan:2023cmz}
A.~Chatrchyan, C.~Er\"oncel, M.~Koschnitzke, and G.~Servant, ``{ALP dark matter with non-periodic potentials: parametric resonance, halo formation and gravitational signatures},'' \href{http://dx.doi.org/10.1088/1475-7516/2023/10/068}{{\em JCAP} {\bfseries 10} (2023) 068}, \href{http://arxiv.org/abs/2305.03756}{{\ttfamily arXiv:2305.03756 [hep-ph]}}.

\bibitem{BICEP:2021xfz}
{\bfseries BICEP, Keck} Collaboration, P.~A.~R. Ade {\em et~al.}, ``{Improved Constraints on Primordial Gravitational Waves using Planck, WMAP, and BICEP/Keck Observations through the 2018 Observing Season},'' \href{http://dx.doi.org/10.1103/PhysRevLett.127.151301}{{\em Phys. Rev. Lett.} {\bfseries 127} no.~15, (2021) 151301}, \href{http://arxiv.org/abs/2110.00483}{{\ttfamily arXiv:2110.00483 [astro-ph.CO]}}.

\bibitem{Cheng:2015oqa}
S.-L. Cheng, W.~Lee, and K.-W. Ng, ``{Numerical study of pseudoscalar inflation with an axion-gauge field coupling},'' \href{http://dx.doi.org/10.1103/PhysRevD.93.063510}{{\em Phys. Rev. D} {\bfseries 93} no.~6, (2016) 063510}, \href{http://arxiv.org/abs/1508.00251}{{\ttfamily arXiv:1508.00251 [astro-ph.CO]}}.

\bibitem{Notari:2016npn}
A.~Notari and K.~Tywoniuk, ``{Dissipative Axial Inflation},'' \href{http://dx.doi.org/10.1088/1475-7516/2016/12/038}{{\em JCAP} {\bfseries 12} (2016) 038}, \href{http://arxiv.org/abs/1608.06223}{{\ttfamily arXiv:1608.06223 [hep-th]}}.

\bibitem{DallAgata:2019yrr}
G.~Dall'Agata, S.~Gonz\'alez-Mart\'\i{}n, A.~Papageorgiou, and M.~Peloso, ``{Warm dark energy},'' \href{http://dx.doi.org/10.1088/1475-7516/2020/08/032}{{\em JCAP} {\bfseries 08} (2020) 032}, \href{http://arxiv.org/abs/1912.09950}{{\ttfamily arXiv:1912.09950 [hep-th]}}.

\bibitem{Garcia-Bellido:2023ser}
J.~Garcia-Bellido, A.~Papageorgiou, M.~Peloso, and L.~Sorbo, ``{A flashing beacon in axion inflation: recurring bursts of gravitational waves in the strong backreaction regime},'' \href{http://dx.doi.org/10.1088/1475-7516/2024/01/034}{{\em JCAP} {\bfseries 01} (2024) 034}, \href{http://arxiv.org/abs/2303.13425}{{\ttfamily arXiv:2303.13425 [astro-ph.CO]}}.

\bibitem{Sobol:2019xls}
O.~O. Sobol, E.~V. Gorbar, and S.~I. Vilchinskii, ``{Backreaction of electromagnetic fields and the Schwinger effect in pseudoscalar inflation magnetogenesis},'' \href{http://dx.doi.org/10.1103/PhysRevD.100.063523}{{\em Phys. Rev. D} {\bfseries 100} no.~6, (2019) 063523}, \href{http://arxiv.org/abs/1907.10443}{{\ttfamily arXiv:1907.10443 [astro-ph.CO]}}.

\bibitem{Domcke:2020zez}
V.~Domcke, V.~Guidetti, Y.~Welling, and A.~Westphal, ``{Resonant backreaction in axion inflation},'' \href{http://dx.doi.org/10.1088/1475-7516/2020/09/009}{{\em JCAP} {\bfseries 09} (2020) 009}, \href{http://arxiv.org/abs/2002.02952}{{\ttfamily arXiv:2002.02952 [astro-ph.CO]}}.

\bibitem{Gorbar:2021rlt}
E.~V. Gorbar, K.~Schmitz, O.~O. Sobol, and S.~I. Vilchinskii, ``{Gauge-field production during axion inflation in the gradient expansion formalism},'' \href{http://dx.doi.org/10.1103/PhysRevD.104.123504}{{\em Phys. Rev. D} {\bfseries 104} no.~12, (2021) 123504}, \href{http://arxiv.org/abs/2109.01651}{{\ttfamily arXiv:2109.01651 [hep-ph]}}.

\bibitem{Domcke:2023tnn}
V.~Domcke, Y.~Ema, and S.~Sandner, ``{Perturbatively including inhomogeneities in axion inflation},'' \href{http://dx.doi.org/10.1088/1475-7516/2024/03/019}{{\em JCAP} {\bfseries 03} (2024) 019}, \href{http://arxiv.org/abs/2310.09186}{{\ttfamily arXiv:2310.09186 [astro-ph.CO]}}.

\bibitem{Dalianis:2018frf}
I.~Dalianis, A.~Kehagias, and G.~Tringas, ``{Primordial black holes from \ensuremath{\alpha}-attractors},'' \href{http://dx.doi.org/10.1088/1475-7516/2019/01/037}{{\em JCAP} {\bfseries 01} (2019) 037}, \href{http://arxiv.org/abs/1805.09483}{{\ttfamily arXiv:1805.09483 [astro-ph.CO]}}.

\bibitem{Witkowski:2022mtg}
L.~T. Witkowski, ``{SIGWfast: a python package for the computation of scalar-induced gravitational wave spectra},'' \href{http://arxiv.org/abs/2209.05296}{{\ttfamily arXiv:2209.05296 [astro-ph.CO]}}.

\bibitem{Caprini_2018}
C.~Caprini and D.~G. Figueroa, ``Cosmological backgrounds of gravitational waves,'' \href{http://dx.doi.org/10.1088/1361-6382/aac608}{{\em Classical and Quantum Gravity} {\bfseries 35} no.~16, (July, 2018) 163001}. \url{http://dx.doi.org/10.1088/1361-6382/aac608}.

\bibitem{Mentasti:2023gmg}
G.~Mentasti, C.~Contaldi, and M.~Peloso, ``{Prospects for detecting anisotropies and polarization of the stochastic gravitational wave background with ground-based detectors},'' \href{http://dx.doi.org/10.1088/1475-7516/2023/08/053}{{\em JCAP} {\bfseries 08} (2023) 053}, \href{http://arxiv.org/abs/2304.06640}{{\ttfamily arXiv:2304.06640 [gr-qc]}}.

\bibitem{Dimastrogiovanni:2024lzj}
E.~Dimastrogiovanni, M.~Fasiello, M.~Michelotti, and O.~\"Ozsoy, ``{A universal constraint on axion non-Abelian dynamics during inflation},'' \href{http://dx.doi.org/10.1088/1475-7516/2025/03/007}{{\em JCAP} {\bfseries 03} (2025) 007}, \href{http://arxiv.org/abs/2405.17411}{{\ttfamily arXiv:2405.17411 [astro-ph.CO]}}.

\bibitem{Ferreira:2015omg}
R.~Z. Ferreira, J.~Ganc, J.~Nore\~na, and M.~S. Sloth, ``{On the validity of the perturbative description of axions during inflation},'' \href{http://dx.doi.org/10.1088/1475-7516/2016/04/039}{{\em JCAP} {\bfseries 04} (2016) 039}, \href{http://arxiv.org/abs/1512.06116}{{\ttfamily arXiv:1512.06116 [astro-ph.CO]}}. [Erratum: JCAP 10, E01 (2016)].

\bibitem{Peloso:2016gqs}
M.~Peloso, L.~Sorbo, and C.~Unal, ``{Rolling axions during inflation: perturbativity and signatures},'' \href{http://dx.doi.org/10.1088/1475-7516/2016/09/001}{{\em JCAP} {\bfseries 09} (2016) 001}, \href{http://arxiv.org/abs/1606.00459}{{\ttfamily arXiv:1606.00459 [astro-ph.CO]}}.

\bibitem{Kitajima:2018zco}
N.~Kitajima, J.~Soda, and Y.~Urakawa, ``{Gravitational wave forest from string axiverse},'' \href{http://dx.doi.org/10.1088/1475-7516/2018/10/008}{{\em JCAP} {\bfseries 10} (2018) 008}, \href{http://arxiv.org/abs/1807.07037}{{\ttfamily arXiv:1807.07037 [astro-ph.CO]}}.

\bibitem{Adshead:2013qp}
P.~Adshead, E.~Martinec, and M.~Wyman, ``{Gauge fields and inflation: Chiral gravitational waves, fluctuations, and the Lyth bound},'' \href{http://dx.doi.org/10.1103/PhysRevD.88.021302}{{\em Phys. Rev.} {\bfseries D88} no.~2, (2013) 021302},
\href{http://arxiv.org/abs/1301.2598}{{\ttfamily arXiv:1301.2598 [hep-th]}}.

\end{thebibliography}\endgroup
\end{document}